\documentclass[10pt,a4paper]{article}
\usepackage[utf8]{inputenc}
\usepackage{setspace}
\usepackage[english]{babel}
\usepackage{pgf,tikz}
\usepackage{lmodern}
\usepackage{setspace}
\usepackage{amsmath}
\usepackage{amsfonts}
\usepackage{amssymb}
\usepackage{amsthm}
\usepackage{mathrsfs}
\usepackage{graphicx}
\usepackage{subcaption}
\usepackage{epsfig}
\usepackage{epstopdf}
\usepackage{lineno}
\usepackage{adjustbox}
\usepackage{chngpage}
\usepackage{array}
\usepackage[compatibility=false]{caption}
\usepackage{titlesec}
\usepackage[a4paper]{geometry}
\usepackage{booktabs}
\usepackage{multirow}
\usepackage{xifthen}
\usepackage{hyperref}
\usepackage{authblk}
\usepackage{listings}
\DeclareMathOperator{\sign}{sign}

\title{A coupled VOF/embedded boundary method to model two-phase flows on arbitrary solid surfaces}
\author[,1]{Mathilde Tavares\thanks{Corresponding author: \texttt{tavaresmathilde92@gmail.com}}}
\author[1]{Christophe Josserand}
\author[4]{Alexandre Limare}
\author[3]{José Mª Lopez-Herrera}
\author[2]{Stéphane Popinet}

\affil[1]{LadHyX, Ecole polytechnique, France}
\affil[2]{Institut Jean Le Rond d'Alembert, CNRS \& Sorbonne University, France}
\affil[3]{Dept. Ing. Aeroespacial y Mec. de Fluidos, ETSI, Universidad de Sevilla, Spain}
\affil[4]{Fluid Mechanics Dept, ArianeGroup, France}

\renewcommand{\v}{u}

\newcommand{\DD}{{\bar{\bar{\cal D}}}}

\newcommand{\nn}{{\bar{n}}}
\renewcommand{\tt}{{\bar{t}}}
\newcommand{\xx}{{\bar{x}}}
\newcommand{\vv}{{\bar{\v}}}

\newcommand{\vvx}{{u_x}}
\newcommand{\vvy}{{u_y}}
\newcommand{\vvz}{{u_z}}

\newcommand{\utau}{{u_{\tau}}}
\newcommand{\un}{{u_{n}}}

\newcommand{\grad}{{\bar{\nabla}}}
\renewcommand{\div}{{\bar{\nabla}\cdot}}

\newcommand{\vol}{{\mathcal{V}}}
\newcommand{\func}{{\mathcal{F}}}

\newcommand{\deff}{\stackrel{\!\text{\tiny def}\!}{\!=\!}}

\begin{document}
\maketitle
\begin{abstract}
  We present an hybrid VOF/embedded boundary method allowing to model two-phase flows in presence of solids with arbitrary shapes. The method relies on the coupling of existing methods: a geometric Volume of fluid (VOF) method to tackle the two-phase flow and an embedded boundary method to sharply resolve arbitrary solid geometries. Coupling these approaches consistently is not trivial and we present in detail a quad/octree spatial discretization for solving the corresponding partial differential equations.
  Modelling contact angle dynamics is a complex physical and numerical problem. We present a Navier-slip boundary condition compatible with the present cut cell method, validated through a Taylor-Couette test case.
  To impose the boundary condition when the fluid-fluid interface intersects a solid surface, a geometrical contact angle approach is developed. Our method is validated for several test cases including the spreading of a droplet on a cylinder, and the equilibrium shape of a droplet on a flat or tilted plane in 2D and 3D. 
  The temporal evolution and convergence of the droplet spreading on a flat plane is also discussed for the moving contact line given the boundary condition (Dirichlet or Navier) used. 
  The ability of our numerical methodology to resolve contact line statics and dynamics for different solid geometries is thus demonstrated.
\end{abstract}

\section{Introduction}
Multiphase flows in complex geometries are present in many environmental contexts and industrial applications, from raindrops impacting on leaves and plants \cite{Amador_2013,Gilet_2015} to droplet deposition on fog harvesting structures \cite{Moncuquet_2022,Protiere_2013, Labb_2019}, from multiphase fluid-structure interactions in powerplant \cite{Merigoux_2016} to 3D printing.
In these situations, three different phases are at least concerned: two fluids, a gas and a liquid in general (any liquid vapor contained in the gas phase, commonly, has a negligible effect on the dynamics); and a solid. This makes these types of problems challenging and difficult to tackle either experimentally or numerically because of the complexity of the flow coupled with generally non-flat, heterogeneous or even porous solid structures.
This can be even more complex when phase change impacts the dynamics, such as in condensation/vaporization or melting/solidification, for instance in steam generators or ice formation.
Numerical modelling tools, combined with theoretical approaches and experimental measurements, have an important role to play for understanding the fundamentals of the fluid dynamical interactions in these applications.
Such three-phase numerical simulations are difficult however, even in the case without phase change. Three main difficulties are combined: (i) liquid-gas interface tracking, (ii) interaction between the liquid-gas and a solid that may be geometrically complex and, finally, (iii) liquid-gas and solid interaction including models of contact line dynamics.\\

(i) When it comes to multiphase flow tracking methods, the challenge is to define sharp interfaces while keeping the accuracy of the interfacial attributes (normal, curvature) and robustness. In past decades, major progress has been achieved and various tracking techniques have been investigated in the literature to handle multiphase flow simulations.

A class of methods for handling fluid-fluid interface on a fixed Eulerian mesh is the front-tracking approach \cite{TryggvasonBunnerEsmaeeliEtAl2001}.
A surface mesh is built on the interface in a Lagrangian way while the Navier-stokes equations are solved on the Eulerian fixed grid.
Front-tracking approaches intrinsically maintain the sharpness of the interface making this approach one of the most accurate methods to track interfaces and deal with capillary effects. However, they involve high coding complexity of dynamical re-meshing process and mesh management in two- and three-dimensional flows. Moreover, they do not automatically handle interface rupture or coalescence.

On the other hand, front-capturing methods capture the presence of the two fluid interface using an auxiliary Eulerian variable. Among them, the level-set method \cite{SussmanOsher1994} is an approach that is very often used. This method uses a signed distance function which is continuous across the interface to locate it. The continuous distribution of phases through the distance function allows accurate estimation of the curvature and normal at the interface.
Despite these advantages, the level-set method generally suffers from a lack of mass conservation.
The Volume Of Fluid (VOF) method \cite{youngs1982} is another front-capturing method for which the interface is located with a ``color function'' which denotes the volume fraction of fluid in a grid cell.
Due in particular to its good mass conservation properties and relative simplicity, the VOF approach is one of the most popular front capturing method even though it requires  specific efforts to transport interfaces without introducing numerical diffusion and to compute an accurate curvature at the interface. Geometric schemes such as SLIC \cite{Noh} and PLIC \cite{youngs1982} have been developed to give a sharp interface representation.
Various authors \cite{Sussman2000,Desjardins,Lechenadec2} have proposed a combination of both level-set and VOF methods to give a simpler representation of the interface thanks to the level-set method while ensuring natural mass conservation with the VOF approach.
More recent developments \cite{Compere,lv-zou-reeve-zhao-2012,DENNER2014,IVEY2017,XIE2017} have been focused on unstructured grids for VOF and level-set methods.\\

(ii) Coupling the liquid-gas system with an arbitrary solid  generally requires computational tools to describe the solid boundary in addition to the approach selected to model the multiphase flows.
Non-body-conforming methods are a class of methods using a Cartesian grid on which the fluid equations of motion (Navier--Stokes) are solved whereas the solid bodies are tracked using a surface mesh independent from the fixed grid.
Here, the Cartesian grid does not conform to the solid body, thus it is necessary to modify the equations in the vicinity of the solid boundary to account for its presence.
For these approaches, the main difficulties arise in the numerical cells where the three phases are present and for which no simple numerical reconstruction is available, in contrast with two-phase flows~\cite{Scardovellietzaleski1999}.
Body conforming methods use this latter principle, fitting the grid to the solid boundary. They might appear appealing at first because they give a sharp description of the solid boundary while fullfilling all the boundary conditions, possibly with a high-order of accuracy.
However, the lack of flexibility in handling complex geometries, the computational overhead associated with the remeshing process and the additional equations for the mesh motion make them less attractive. Non-body-conforming methods such as the popular Immersed Boundary Methods (IBMs) provide great advantages over traditional body conforming methods since they allow simple grid generation and discretization of the Navier-Stokes equations while requiring less memory. IBMs were originally proposed by \cite{Peskin_1972} for cardiac flow modelling and are used now for many different applications, mostly for fluid-structure interactions. IBMs can be classified into two groups:
\begin{itemize}
\item Continuous forcing (CF) approaches \cite{Peskin_1972} use a continuous forcing source term based on a regularized Dirac function in the momentum equation to enforce no-slip boundary condition at the solid frontier. CF methods are very attractive for flows with immersed elastic boundaries (biology, interfacial flows). However, in the rigid limit, forcing terms used in these approaches behave less well with for instance problems of numerical accuracy and stability. They are notoriously non-sharp since the regularized Dirac function kernel requires a compact support on Eulerian grids and therefore spreads the solid boundary over a few cells.
\item With discrete forcing approaches (DF), the boundary conditions are accounted for at the level of the discretized momentum equations. Since the forcing procedure is directly linked to the discretization method used, DF approaches are not as straightforward to implement as CF methods. However, the discrete nature of the forcing in DF approaches allow to generate sharp immersed boundaries and do not require any stability constraints in the solid body representation. DF approaches can be classified in two categories. Indirect boundary condition imposition on one side and direct boundary condition imposition on the other side with the ghost-cells finite difference method and the cut cell finite volume approach.
\end{itemize}
A review of the large variety of IBMs is presented in \cite{Mittal_2005}.
Due to the sharp representation of the solid boundary in DF approaches, the cut-cell or embedded boundary method \cite{Clarke_1986,Udaykumar1996,JOHANSEN1998,Popinet2003,SCHWARTZ2006,Ghigo}, which is second-order accurate and conservative, is used in this work, and a more detailed description will be given in the following sections.

(iii) When coupling solid and multiphase flows, different issues have to be considered at their interface, corresponding to boundary conditions that have to be correctly implemented:
when the two fluids are in contact with the solid (wetting or dewetting problems), the coupling problem lies in the correct numerical modelling of the contact angle that the liquid/gas interface adopts at the solid frontier. Microscopic physico-chemical interactions between molecules of the two immiscible fluids and the solid drive the contact line behavior.
At the macroscopic scale, one can see this contact line as an apparent contact angle between the liquid/gas interface and the solid substrate. Multiple studies throughout the literature have been carried out to give a modelling of this macroscopic apparent contact angle with either static or dynamical value but generally using simple solid geometries \cite{Afkhami_2009,DUPONT2010,Legendre_2015}.
In the last decade, several attempts have been made for more complex geometries using generally IBMs coupled with either VOF, level-set or front-tracking to model the multiphase flow.
A diffuse interface immersed boundary method for the simulation of flows with moving contact lines on curved substrates was proposed by \cite{Liu_2015}, where they considered only 2D problems.
The authors of \cite{PATEL2017} used in their work a VOF/IBM coupling with an implicit DF approach to simulate 3D multiphase flow and solid interaction with contact line dynamics. A VOF coupled to a DF ghost-cells/IBM to enforce contact angle dynamics on 2D arbitrarily solids has been proposed by \cite{O_Brien_2018}.
Another immersed boundary based contact angle method coupled with VOF was developed by \cite{G_hl_2018} to handle complex surfaces of mixed wettabilities with a focus on contact angle models.
More recently, \cite{asghar2023} proposed a wetting benchmark using an unstructured VOF method including a contact line model, allowing to deal with complex solid geometries.

In this article, we present an hybrid model coupling a geometric volume of fluid (VOF) method representing the liquid-gas interface with a cut-cell embedded boundary method accounting for the solid body. The resulting scheme maintains a sharp interface and solid boundary representation, while ensuring mass conservation.
While both methods are nowadays classical and have been extensively validated for two-phase flows on one side and solid on the other side, their coupling remains scarce \cite{Udaykumar1996}.
Different boundary conditions for the velocity fields and numerical schemes are addressed in order to be able to model the contact line dynamics which are crucial in such three-phases flows on complex geometries.

This paper is structured as follows: the second section is dedicated to the description of the governing equations with a detailed description of the embedded boundary method and the coupled VOF/embedded boundary method. The contact angle model developed in this work is also presented in static and dynamical situations.
In the third section, validation test cases are proposed to show the ability of our approach to accurately deal with three-phase problems involving a triple line in two- or three-dimensional test cases such as a droplet wetting a cylinder or a flattened droplet on an horizontal or inclined embedded plane.

\section{Governing equations}\label{sec:sec1}
We consider two incompressible, Newtonian and non-miscible fluids and a non-deformable solid so that the coupling between the fluids and the solid intervenes only through the boundary conditions.
As a first step, we assume that there is no mass and heat transfer at the fluid-fluid interface and that the surface tension between the two fluids $\sigma$ is constant. The two fluids will be denoted $g$ and $l$ for gas and liquid, with densities and dynamical viscosities $\rho_g$ and $\rho_l$, $\mu_g$ and $\mu_l$ respectively.

The dynamics of the two fluids are therefore governed by the unsteady incompressible, variable density, Navier-Stokes equations which can be written, in the classical one-fluid formulation \cite{Kataoka}: 
\begin{subequations}
\begin{equation}
 \frac{\partial{\rho} }{\partial{t}}   + \div (\rho \vv) =0,
\label{eq:variable-density}
\end{equation}
\begin{equation}
\div \vv = 0,
\label{eq:continuity}
\end{equation}
\begin{equation}
\rho\left(\frac{\partial{\vv}}{\partial{t}} + \div(\vv\otimes\vv)\right) = -\grad P + \div (2\mu\DD) + \rho \bar{g} +  \bar{F}_{\sigma}
\label{eq:momentum}
\end{equation}
\label{eq:Navier-stokes}
\end{subequations}
where $\vv\equiv\vv(\xx,t)$ is the flow velocity field, $P \equiv P(\xx,t)$ stands for the pressure, $\DD = (\grad \vv +{\grad \vv}^T)/2 $ is the rate of deformation tensor, $\rho\equiv\rho(\xx,t)$ the one fluid density, $\mu\equiv\mu(\xx,t)$ the one fluid viscosity, $\bar{F}_{\sigma}$ the surface tension force and $\bar{g}$ is the acceleration due to the gravity.
The surface tension force is modeled using the continuum surface force \cite{Brackbill} (CSF) so that $\bar{F}_{\sigma} = \sigma \kappa {\nn\delta}_{\Gamma_F}$ with $\sigma$ the surface tension, $\kappa$ the curvature, $\nn$ the normal unit vector to the interface and $\delta_{\Gamma_F}$ the Dirac distribution function defined by $\delta_{\Gamma}= \delta(\xx - \xx_{\Gamma_F})$ indicating that it is equal to zero everywhere  except on the interface \textit{i.e} $\xx = \xx_{\Gamma_F}$.\\
Within the one fluid formalism, a discontinuous indicator function $\chi$ is used to locate the interface:
\begin{equation}
\chi(\xx,t) = \left\{
\begin{array}{cc}
0 &\mbox{in\, the \, gas}\\
1 &\mbox{in\, the \, liquid}
\end{array}
\right .
\label{eq:chi_k}
\end{equation}
The evolution of the interface is then simply given by the following advection equation:
\begin{equation}
\frac{\partial{\chi}}{\partial{t}} + \vv. \grad \chi = 0 \label{eq:volume-fraction}
\end{equation}
The density $\rho$ and the viscosity $\mu$ are fields that also depend on space and time through the relation:
\begin{eqnarray}
\rho(\xx,t) &=& \chi \rho_l + (1 - \chi)\rho_g \label{eq:local-rho} \\
\mu(\xx,t) &=& \chi \mu_l + (1 - \chi)\mu_g \label{eq:local-mu}
\end{eqnarray}
Boundary conditions (BC) between the fluids and the solids have now to be added to this set of equations: indeed the one fluid formulation already provides by construction the correct boundary conditions between the two fluids, through the CSF term.
At the solid/fluids interfaces, two different types of boundary conditions need to be imposed: on the velocity fields on the one hand, on the indicator function $\chi$ on the other hand.
As stated above, we consider the solid as a non-deformable body so that its velocity can be simply described by the classical solid body motion $\bar{U}_s(x,t)=\bar{U}_0+ \bar{\Omega} \times \xx $. Therefore, since we have viscous fluids, the usual boundary conditions for the fluid velocity fields at the solid interface reads:
$$ \bar{u}(\xx,t)= \bar{U}_s(\xx,t) \,\, {\rm for} \;\; \xx  \in \xx_{\Gamma_F}.$$
This boundary condition imposes in fact both the non-penetration of the fluids inside the solid (the continuity of the normal velocity at the interface) and the no-slip boundary condition for the tangential velocity fields because of the viscous stress. This latter condition is often numerically relaxed by introducing a slip at the solid boundary whose origin can be microscopic (mean free path) and/or geometric (effective roughness of the solid surface), leading to the
so-called Navier condition:
\begin{equation}\label{eq:Navier}
\lambda \frac{\partial \vv_\tau}{\partial n} + (\vv_\tau-\bar{U}_{s,\tau})=0,
\end{equation}
where $\lambda$ is the slip length, while $\vv_\tau$ is the projection of the velocity tangentially to the solid (with $\bar{U}_{s,\tau}$ the tangential component of the body motion).
Similarly, for the interface on the solid, we need to impose the correct boundary condition for the contact line leading eventually to a condition on the characteristic function $\chi$. This corresponds to imposing the contact angle between the interface and the solid surface at the contact line. While the contact angle $\theta_s$ can be known for static configurations, where the Young-Laplace relation holds, it is much more complex to determine when the contact line is moving, a problem which has been much debated in the literature~~\cite{deGennes1985,RMPBonn09}. For the rest of the present paper, we will simply assume that this dynamical contact angle $\theta_d$ is given by a model, typically a function of the velocity in the vicinity of the contact line. For validating the numerical scheme we will use the simplest model for the moving contact line consisting in imposing a constant contact angle coupled with a Navier-slip condition for the velocity.

\section{Description of the numerical method}\label{sec:sec2}
The numerical scheme developed here is based on the open source library for partial-differential equations, Basilisk \cite{Popinet2003, Popinet2009, POPINET2015}. A detailed description of Basilisk can be found at (\url{http://basilisk.fr}). Consequently, we will only give a brief description of the elements when they are not specific to the coupled VOF/embedded boundary implementations. An effort is devoted to the description of the VOF approach used on one side and the embedded boundary on the other, to fully understand the features of the coupled VOF/embedded boundary method. The section is divided into four main parts: (i) the temporal discretization focusing on numerical solution of the incompressible Navier-Stokes equations, (ii) the VOF approach used to deal with the dynamics of deformable fluid-fluid interface, (iii) the embedded boundary method handling the presence of solids, (iv) the coupled VOF/embedded boundary method with its main features, namely when a contact angle boundary condition is required to ensure the triple point/line definition at the intersection between the fluid-fluid interface and the solid.

In order to discretize the equations, the volume fraction $F$ corresponding to the average value of $\chi$ in each computational cell $\Omega_F$ of characteristic volume $\vol$ is introduced:
\begin{equation}
F(\xx,t) = \frac{1}{\vol}\int_{\Omega_F} {\chi(\xx,t)dV}
\label{eq:fraction_volumique}
\end{equation}
Then, the evolution of the interface is given by the following advection equation:
\begin{equation}
\frac{\partial{F}}{\partial{t}} + \vv. \grad F = 0 \label{eq:volume-fraction}
\end{equation}
The local density $\rho$ and viscosity $\mu$ are defined from the local volume fraction $F$ using the standard arithmetic mean:
\begin{eqnarray}
\rho(\xx,t) &=& F\rho_l + (1 - F)\rho_g \label{eq:local-rho} \\
\mu(\xx,t) &=& F \mu_l + (1 - F)\mu_g \label{eq:local-mu}
\end{eqnarray}

\subsection{Temporal discretization}\label{sec:sec2subsec1}
The discretization of the Navier–Stokes equations \eqref{eq:Navier-stokes} is summarized here. The incompressible Navier-Stokes equations are approximated using a a time-staggered approximate projection method on a Cartesian grid. The advection term is discretized with the explicit and conservative Bell--Colella--Glaz second-order unsplit upwind scheme \cite{Bell1989}. For the discretization of the viscous diffusion term, a second-order Crank--Nicholson fully-implicit scheme is used. Spatial discretization is achieved using a quadtree (octree in three dimensions) adaptive mesh refinement on collocated grids.
The adaptive mesh technique is based on the wavelet decomposition \cite{POPINET2015, vanhooft2018} of the variables such as the velocity field or the volume fraction and is efficient and critical to the success of some simulations as it allows high resolution only where needed, therefore reducing the cost of calculation.

The following fractional step method is used:
\begin{itemize}
  \item The Navier-stokes equations \eqref{eq:continuity}-\eqref{eq:momentum} are first discretized as
    \begin{eqnarray}
   \rho^{n+\frac{1}{2}}_c \left(\frac{\vv^{\ast}_c - \vv^{n}_c}{\Delta t} + \vv^{n+\frac{1}{2}}_c.\grad\vv^{n+\frac{1}{2}}_c\right) &=&  [-\grad P^n + \bar{a}^{n}]_{f\rightarrow c} \\
    \vv^{\ast}_c &=& \vv^{\ast}_c +  \Delta t [-\grad P^n + \bar{a}^{n}]_{f\rightarrow c} \\
    \rho^{n+\frac{1}{2}}_c\left(\frac{\vv^{\ast\ast}_c - \vv^{\ast}_c}{\Delta t}\right) &= & \div (2\mu^{n+\frac{1}{2}}_f\DD^{\ast\ast}) \\
    \vv^{\ast\ast}_c &=& \vv^{\ast\ast}_c -  \Delta t [-\grad P^n + \bar{a}^{n}]_{f\rightarrow c} \\
    \vv^{n+1}_f &=& \vv^{\ast\ast}_{c\rightarrow f} + \Delta t \bar{a}^{n+1}_f \label{eq:uf_temp}
    \end{eqnarray}
with $\Delta t\deff  t^{n+1} - t^{n}$ the time step. The subscript $_c$ and $_f$ refers to the cell center and cell face respectively. Equation \eqref{eq:uf_temp} gives a temporary velocity $\vv^{n+1}_f$ obtained by an arithmetic averaging from the centered velocity field $(_{c\rightarrow f})$ to which the acceleration field $\bar{a}^{n+1}_f$ is added. This acceleration term is simply due to the surface tension forces and the gravity, by considering ($\bar{a} = \sigma \kappa {\nn\delta}_{\Gamma_F}+ \rho\bar{g}$). To ensure a consistent discretization, the acceleration term is defined on the faces $_f$ like the pressure gradient, in order to minimize the parasitic currents close to the interface, according to the well-balanced surface tension model \cite{Popinet2009,Popinet2018}. The combined centered acceleration and pressure gradient is also obtained by arithmetic averaging $(_{f\rightarrow c})$.
\item To ensure incompressibility, the temporary velocity \eqref{eq:uf_temp} is projected so that:
\begin{equation}
\div \vv^{n+1}_f = \div\vv^{\ast\ast}_{c\rightarrow f} + \div \left[  \Delta t ( -\grad P^n + \bar{a}^{n+1} )\right]_f
\label{eq:projection}
\end{equation}
with $\div \vv^{n+1}_f = 0$. The pressure $P^{n+1}$ is then obtained using a multigrid iterative method to solve Eq.~\eqref{eq:projection}. We get the divergence free face velocity $\vv^{n+1}_f$ by substituting $P^{n+1}$ in Eq.~\eqref{eq:projection}.
\item The final centered velocity Eq.~\eqref{eq:uc_final} is obtained using the combined pressure gradient and acceleration interpolated from faces to cells.
\begin{equation}
\vv^{n+1}_c = \vv^{\ast\ast}_c +  \Delta t [-\grad P^n + \bar{a}^{n}]_{f\rightarrow c} \\
\label{eq:uc_final}
\end{equation}
\end{itemize}
The cell-centered velocity thus verifies $\div \vv^{n+1}_c\approx 0$ while the face-centered velocity satisfies $\div \vv^{n+1}_f= 0$, hence the name ``approximate projection method" \cite{Popinet2003}.

\subsection{Interface advection (VOF)}\label{sec:sec2subsec3}
The geometric VOF PLIC method on an Eulerian fixed grid is used, because of its accuracy to maintain a sharp interface between the fluids while ensuring mass conservation. \\
The integration of the hyperbolic advection equation Eq.~\eqref{eq:volume-fraction} is done in two steps. First, the explicit interface is locally reconstructed from the volume fraction and then the advection is performed using a split advection method.

\subsubsection{Geometric reconstruction of the interface} \label{sec:sec2subsubsec1}
Consider a discrete representation of the liquid-gas system separated by the interface $\Gamma_F$, integrated on a regular Cartesian grid, composed of squares (cubes in 3D) cells of size $\Delta$. Each cell is a control volume $\vol$, delimited by the closed contour $\delta\vol$ and associated with a velocity or pressure variable.
The fluid-fluid interface $\Gamma_F$ is represented with a piecewise-linear reconstruction (VOF PLIC), satisfying in each cut cell the following equation for a line (or plane in 3D):
\begin{equation}
\nn_{\Gamma_F}\cdot\xx = \alpha_F
\label{eq:intercept}
\end{equation}
where $\nn_{\Gamma_F}$ is the the local normal to the interface, $\xx$ the vector position and $\alpha_F$ the intercept.
The normal $\nn_{\Gamma_F}$ is estimated using the Mixed-Youngs-Centred (MYC) implementation of \cite{Aulisa_2007}. By ensuring that the fluid volume contained within a plane of normal $\nn_{\Gamma_F}$ is equal to $F$, the intercept $\alpha_F$ can be determined in a unique way, using analytical relations \cite{SCARDOVELLI2000} already implemented in Basilisk: $F = \func (\nn_{\Gamma_F},\alpha_F)$ or  $\alpha_F= \func^{-1} (\nn_{\Gamma_F},F)$.

\subsubsection{VOF advection and geometric flux computation} \label{sec:sec2subsubsec2}
Once the reconstruction step is completed, the advection equation:
\begin{equation}
\frac{\partial{F}}{\partial{t}} + \div(F \vv) = F\div\vv
\label{eq:volume-fraction-1d}
\end{equation}
is integrated using dimension-splitting. The method proposed by \cite{WEYMOUTH2010} is used to discretize the equation \eqref{eq:volume-fraction-1d} which is solved along each dimension successively, with a fractional step method in order to guarantee exact mass conservation for the multi-dimensional advection scheme as long as the velocity field is divergence free.
The left-hand-side term is computed using the net fluxes per direction whereas the second term represents the dilatation required to compensate the non-divergence-free flow advection per direction.
With an update of the interface from $t^{n-\frac{1}{2}}$ to $t^{n+\frac{1}{2}}$ (it is assumed here that the interface lags the velocity/pressure fields by half a time step), we get:
\begin{eqnarray}
\frac{F^{\ast} - F^{n-\frac{1}{2}}}{\Delta t} + \frac{G^{n}[0,0] - G^{n}[0,1]}{\Delta}  = C_c^{n-\frac{1}{2}}\frac{\vvx^{n}_f[0,0] - \vvx^{n}_f[0,1]}{\Delta} \\
\frac{F^{n+\frac{1}{2}} - F^{\ast}}{\Delta t} + \frac{H^{n}[0,0] - H^{n}[1,0]}{\Delta}  = C_c^{n-\frac{1}{2}}\frac{\vvy^{n}_f[0,0] - \vvy^{n}_f[0,1]}{\Delta}
\end{eqnarray}
where $G$ and $H$ are the face fluxes estimated respectively by $F {\vvx}_f$ and $F {\vvy}_f$.
To ensure exact volume conservation, the centered volume fraction field $C_c^{n-\frac{1}{2}}$ is set to:
\begin{equation}
C_c^{n-\frac{1}{2}} = \left\{
\begin{array}{ccc}
1 &\mbox{if}& F^{n-\frac{1}{2}}>0\\
0 &\mbox{else.}&
\end{array}
\right.
\label{eq:c_c}
\end{equation}
As shown in Figure \ref{fig:flux-fluid-cell}, the fluxes $G$ and $H$ are computed using the geometry of the reconstructed interface \cite{Noh,Debar1974}.  In this illustration, the flux $G$, corresponding to the horizontal face velocity $F \vvx_f$ and delimited by the dotted line is estimated by the dark red area.
This area can be computed using an approach similar to that in the previous section.
Once the volume fraction $F^{n+\frac{1}{2}}$ is updated, the local properties $\rho^{n+\frac{1}{2}}$ and $\mu^{n+\frac{1}{2}}$ can be evaluated using Eqs.~\eqref{eq:local-rho}--\eqref{eq:local-mu}.

\subsection{Computing the surface tension force}\label{sec:sec2subsec4}
In the framework of VOF methods, the continuum surface force model (CSF)  \cite{Brackbill} appears to be the natural way to compute the surface tension term in the momentum equation.
\begin{equation}
\bar{F}_{\sigma} = \sigma \kappa {\nn\delta}_{\Gamma_F}
\label{eq:CSF}
\end{equation}
As mentioned previously (see Sec.~\ref{sec:sec2subsec1}), the surface tension force is computed here using the balanced-force calculation method \cite{Popinet2009,Popinet2018}. Both pressure gradient and surface tension terms are defined on cell faces to ensure a consistent discretization and eventually reach an exact balance between these two terms.
Computing the curvature accurately is more difficult and is done here using the generalized height-function approach of \cite{Popinet2009}.

\subsection{The embedded boundary or cut cell method}\label{sec:sec2subsec2}
The Cartesian grid embedded boundary method, also known as cut cell method, used in this work has already been implemented in Basilisk and used for moving boundaries \cite{Ghigo} and solidification modelling~\cite{Limare}. We give here the general lines of this method, originally designed by \cite{JOHANSEN1998,SCHWARTZ2006}.

\subsubsection{Solid-fluid interface description}
\begin{figure}[!h]
\begin{subfigure}[b]{0.45\textwidth}
\includegraphics[trim={6cm 5cm 2cm 4.5cm},clip,width=1.6\textwidth]{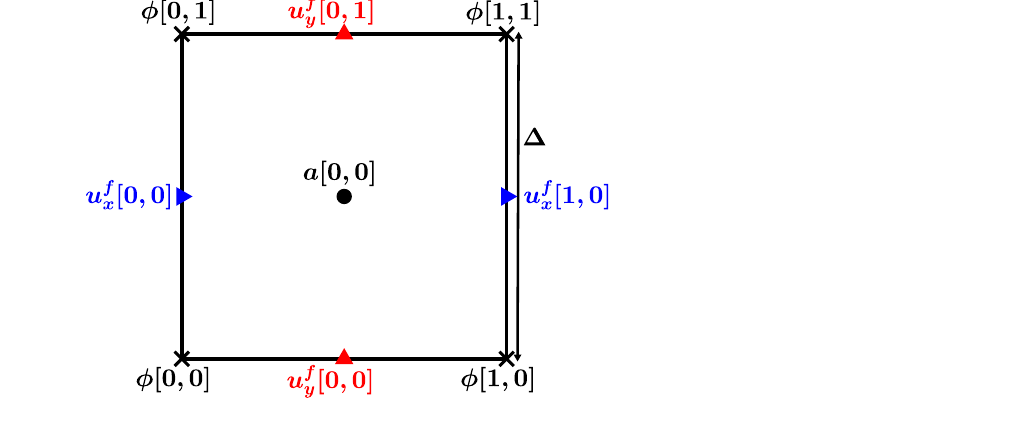}
\caption{}
\label{fig:vertex-cut-cell}
\end{subfigure}
\begin{subfigure}[b]{0.45\textwidth}
\includegraphics[trim={7cm 6.9cm 2cm 4cm},clip,width=1.62\textwidth]{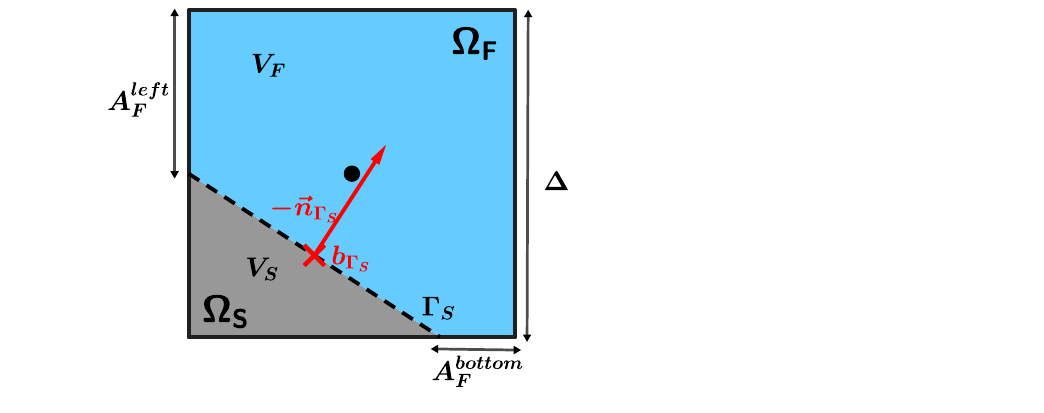}
\caption{}
\label{fig:embedded-cut-cell}
\end{subfigure}
\caption{Stencils used in Basilisk and their respective indexing on a 2D grid: (a) Vertex $\Phi$ and face velocity $\vv_f$ stencils, (b) Cartesian grid cut by the discrete rigid boundary to form irregular control volumes}
\end{figure}

A discrete representation of a solid domain $\Omega_s$ with boundary $\Gamma_s$ is embedded in a regular Cartesian grid, composed of square (cubic in 3D) cells of length $\Delta$ as shown in Fig.~\ref{fig:embedded-cut-cell}.
Each cell is a control volume $\vol$, delimited by the closed contour $\delta\vol$ consisting of edges $f_d$ (faces in 3D).
The index $d$ refers to the direction of the cell so that $d$ is the left, bottom, right or upper direction for a 2D cell.
The reconstruction of the solid interface $\Gamma_S$ is done using a piecewise linear reconstruction thus satisfying in each cut cell the following equation for a line (a plane in 3D):
\begin{equation}
\nn_{\Gamma_S}\cdot\xx = \alpha_S
\label{eq:intercept_S}
\end{equation}
where $\nn_{\Gamma_S}$ is the the local normal to the interface, $\xx$ the position vector and $\alpha_S$ the intercept. The subscript $_F$,$_S$ refers respectively to the fluid and solid domain.
We consider firstly the simplified case where the solid is embedded in a single-phase flow. The coupling with the two-phase flow is detailed in the section hereafter.
The intersection of the discrete solid boundary with the Cartesian cells, forms irregular control volumes in each cell cut by the boundary and separates the control volume $\vol$ in a fluid and a solid subdomain of respective volumes $V_F$ and $V_S$.
A new fluid volume fraction $C$ is defined and used to characterize these irregular volumes so that $C=0$ in solid cells and $C=1$ in liquid cells whereas $0<C<1$ in cut cells.
The effective volume occupied by the fluid $V_F$ in the cell is given by:
\begin{equation}
V_F=C\Delta^D \label{eq:solid-volume-fraction}
\end{equation}
where $D$ is the dimension of the problem. The complementary solid volume is therefore $V_F=(1-C)\Delta^D$.
Similarly a fluid face fraction $0<f_F^d<1$ is defined for each face $d$ so that the effective area occupied by the fluid on a cut face is:
\begin{equation}
A_F^d=f_F^d\Delta^{D-1} \label{eq:solid-face-fraction}
\end{equation}
and the solid area is simply given by $A_S^d=f_S^d\Delta^{D-1} = (1-f_F^d)\Delta^{D-1}$.\\
With this integral description of the discrete rigid boundary, it is possible to represent arbitrary solid shapes whose size is greater than the cell size $\Delta$.

\subsubsection{Embedded fractions in a cut cell}
The geometric properties including the normal, the face and the volume fraction of the interface are computed using a signed distance function $\phi(\xx -\xx_{\Gamma_S})$ sampled on the vertices of the Cartesian grid cell (see Fig.~\ref{fig:embedded-cut-cell}). This signed distance function represents the solid-fluid interface $\Gamma_S$ and is taken by convention positive ``inside" the interface when $\xx < \xx_{\Gamma_S}$ and negative outside when $\xx > \xx_{\Gamma_S}$.
Figure \ref{fig:vertex-cut-cell} shows the description of a Cartesian square (cubic in 3D) cell of size unity centered on the origin.
We give here some elements of the interface characteristics calculation for a 2D cell. Consider for instance the bottom left and right vertices, respectively $\phi[0,0]$ and $\phi[0,1]$ to evaluate the bottom line fraction $f^d = f_F^{y}$.
\begin{itemize}
\item For each edge of an interfacial cell (\textit{i.e} $\phi[0,0]\times\phi[0,1] < 0$), the bottom line fraction for instance $f_F^y$ is given by:
\begin{equation}
{f_F^{y}} = \frac{1 - \sign(\phi[0,0])}{2} + \sign(\phi[0,0])\frac{\phi[0,0]}{\phi[0,0] -\phi[0,1]}
\label{eq:line-fraction}
\end{equation}
\item The normal $\nn_{\Gamma_S}$ is found by expressing the circulation along the closed boundary $\delta\vol$ of the volume control $\vol$:
\begin{equation}
\oint_{\delta\vol} \nn dl \Leftrightarrow \sum_{d=l,b,r,u} f^d \nn^d + \hat{n}_{\Gamma_S} = 0 ~\text{with}~ \nn_{\Gamma_S} = \frac{\hat{n}_{\Gamma_S}}{\left\|\hat{n}_{\Gamma_S}\right\|}
\end{equation}
where $\nn^d$ is the outward unit normal of each face $f^d$ and $\hat{n}_{\Gamma_S}$ the non-unit outward normal to the interface ${\Gamma_S}$.
\begin{equation}
 \hat{n}_{\Gamma_S} = \left(
\begin{array}{cc}
f_F^{x}[0,0]-f_F^{x}[1,0] \\
f_F^{y}[0,0]-f_F^{y}[0,1]
\end{array}
\right)
\label{eq:local-normal}
\end{equation}
\item For each edge, if $0<f^d<1$, then $\alpha_S^d = \nn_{\Gamma_S}\cdot\xx^d$. For the bottom edge, the coordinates of $\xx_d$ are for example:
\begin{equation}
 \xx^d = \left(
\begin{array}{c}
\sign(\phi[0,0])f_F^{y}[0,0]-0.5 \\
-0.5
\end{array}
\right)
\label{eq:local-position}
\end{equation}
The final intercept $\alpha_S$ is  the average of all face intercepts $\alpha^S_d$ for all intersections.
\item Given $\nn_{\Gamma_S}$ and $\alpha_S$, the interface is fully described and the volume fraction can be calculated using the predefined functions \cite{SCARDOVELLI2000} mentioned above so that $C = \func (\nn_{\Gamma_S},\alpha_S)$
\end{itemize}
The generalization to 3D is made quite straightforward by considering the faces of the cell instead of the edges. The whole procedure described above to get the volume fraction $C$ in 2D is now used to obtain the face fraction $f_d$. The rest of the procedure consists in getting the normal, the intercept and finally the volume fraction with some adaptations to account for the third dimension.
The complete procedure of the geometric calculation of the interface properties is detailed in \url{http://basilisk.fr/src/fractions.h}.
From the face fractions, we are also able to define the position of the centroid $\bar{b}_{\Gamma_S}$ in a cut cell:
\begin{equation}
\bar{b}_{\Gamma_S} = \frac{1}{|A_S^{\Gamma}|}\int_{\Gamma_S}\bar{b}dA_S^{\Gamma},
\label{eq:centroid}
\end{equation}
which is an important quantity for approximating differential operators on interfacial cells.

\subsubsection{Discrete operators in a cut cell}
Once the embedded fractions and the centroids are computed, it is possible to write a conservative finite volume approximation of discrete operators in the cut cell such as the divergence of a quantity $\bar{\Phi}$ (typically needed to compute the non linear advection term or the viscous term Eq.~\eqref{eq:momentum} in the fluid side for instance):
\begin{multline}
\div \bar{\Phi} \approx \frac{1}{V_F}\int_{\vol}\div\bar{\Phi}dV_F =\frac{1}{V_F}\int_{\delta\vol}\bar{\Phi}.\nn_d dA =\left(\sum_{d=l,b,r,u} f_F^d \bar{\Phi}.\nn_d\right) + \bar{\Phi}_{\Gamma_S}.\nn_{\Gamma_S}
\label{eq:div-embed}
\end{multline}
where $\nn_d$ is the outward unit normal vector to the face $d$ of the closed boundary $\delta\vol$ within the cell. The challenge is thus to estimate accurately such operator taking into account the correct boundary conditions at the solid/fluid frontier. 
The quantity $\bar{\Phi}_{\Gamma_S}$ is discretized at the centroid $b_{\Gamma_s}$ as well as $\bar{\Phi}$ which is taken at the center of the full or ``regular" Cartesian face $d$ rather than at the center of the irregular fluid face, even when the center of the Cartesian cell is outside the fluid domain. Here, it is assumed that $\bar{\Phi}$ is sufficiently smooth to be extended to the cut cell ($0<C<1$), even if its center is included in the solid part of the cell. This leads to a conservative and at least second-order accurate, finite-volume methodology.

The detailed algorithms and demonstration of 2nd order accuracy of the embedded boundary method can be found in \cite{JOHANSEN1998,Popinet2003}. The interest in using embedded boundary on Cartesian grids over unstructured grid methods or other immersed boundary methods is the generation of simple grids and data structures in 2D or 3D and the straightforward extension to quad/octree meshes.
Complete details of how this method is implemented in Basilisk are available in \url{http://basilisk.fr/src/embed.h}.

Finally, a crucial issue with cut-cell methods is to implement accurately the correct boundary conditions at the solid boundaries for the velocity fields and the contact angle in the case of multiphase flows. While the non-penetrating condition for the velocity field does not require additional treatment for the mass conservation equation, the conditions for the velocity field needs to be taken into account when estimating the viscous term in the Navier-Stokes equations. We therefore investigate firstly how to impose any boundary condition for the velocity in the next section. 

\subsection{Viscous flux through the discrete rigid boundary in a cut cell}
For computing the viscous term in the Navier-Stokes equations, we need, in our formulation, to access the gradients of the velocities in order to evaluate the fluxes associated with the viscous stress tensor.
Considering an incompressible flow and constant viscosity, the viscous term of the Navier-Stokes equations Eq.~\eqref{eq:momentum} can be further simplified so that:
\begin{equation}
\div (2\mu (\grad \vv +{\grad \vv}^T)/2) = \grad^2 (\mu \vv)
\label{eq:simlified-viscous-term}
\end{equation}
The right hand term of the relation \eqref{eq:simlified-viscous-term} then gives decoupled scalar Poisson equations for each velocity component.
To compute the viscous flux through the discrete rigid boundary ${\Gamma_S}$ in a cut cell, we then need to write the conservative finite-volume approximation for each velocity component
\begin{equation}
\int_{\Omega_S} \grad^2 (\mu \vv)_x = \int_{\Gamma_S} \mu \grad \vvx\cdot\nn_{\Gamma_S} dS,
\label{eq:viscous-flux}
\end{equation}
taking the horizontal component of the velocity for instance. Here, $dS$ is the elementary boundary surface and $\grad \vvx\cdot\nn_{\Gamma_S}$ is the embedded face gradient that can also be written $\grad \vvx|_{\Gamma_S}$.
The gradient must be determined at the centroid of each face $f^d$ and also at the centroid $\bar{b}_{\Gamma_S}$ (see \ref{eq:div-embed}) of the cut cell interface $\Gamma_S$ (see fig.~\ref{fig:embedded-cut-cell}) .
Since the calculation of the flux through the face $f^d$ is detailed in \cite{Ghigo}, we focus here solely on the gradient calculation over the discrete rigid boundary $\Gamma_S$ of centroid $\bar{b}_{\Gamma_S}$, which is crucial for the accuracy of the method.
We consider both Dirichlet and Navier boundary conditions on the velocity.

\subsubsection{Dirichlet boundary condition}
If a Dirichlet boundary condition  $\vvx_{\Gamma}$ or $\vvy_{\Gamma}$ ($\vvz_{\Gamma}$ in 3D) is required for the velocity on the solid embedded boundary, the following second-order discretization is used to compute the embedded face gradient of the scalar velocity component. Following the methodology presented in \cite{JOHANSEN1998}, we write for the horizontal component:
\begin{equation}
\grad \vvx|_{\Gamma_S}=\frac{1}{d_2-d_1}\left[\frac{d_2}{d_1}(\vvx_{\Gamma}-{\vvx}^{I_1})-\frac{d_1}{d_2}(\vvx_{\Gamma}-{\vvx}^{I_2})\right]
\label{eq:dirichlet-gradient}
\end{equation}
\begin{figure}[!h]
 \centering
  \includegraphics[trim={4cm 4cm 4cm 2cm},clip,width=0.6\linewidth]{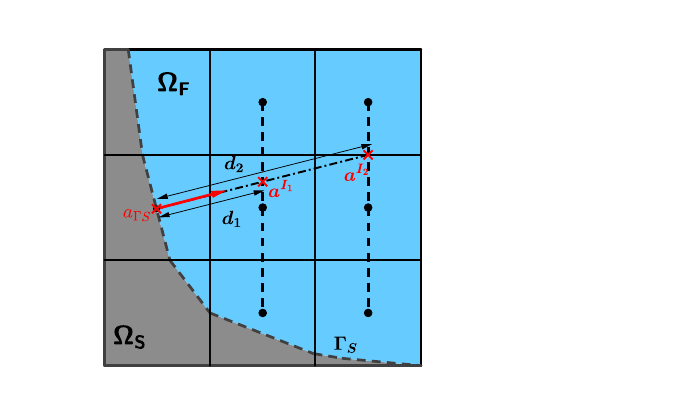}
\caption{Gradient calculation in the embedded boundary method depending on the normal orientation $\nn_{\Gamma_S}$.}
\label{fig:gradient-dirichlet}
\end{figure}
The gradient $\grad \vvx|_{\Gamma_S}$ or $(\grad \vvx\cdot\nn_{\Gamma_S})$ is evaluated at the centroid $b_{\Gamma_S}$ of the cut-cell (see Fig.~\ref{fig:gradient-dirichlet}). The second-order accuracy of the gradient is achieved by computing $\vvx^{I_1}$ and $\vvx^{I_2}$ with a third-order quadratic (biquadratic in 3D) interpolation along the direction orthogonal to the direction of the normal $\nn_{\Gamma_s}$ at distances $d_1$ and $d_2$ from the centroid $b_{\Gamma_s}$ (see \url{http://basilisk.fr/src/embed.h#dirichlet-boundary-condition} for more details).
\\
\subsubsection{Neumann boundary condition}
To impose Neumann boundary condition $s_{\Gamma}$ on the velocity component $\vvx$, the gradient $\grad \vvx|_{\Gamma_S}$ is thus defined precisely at the interface and we simply have to take for evaluating the viscous flux (\ref{eq:viscous-flux}) this value:
\begin{equation}
 \grad \vvx|_{\Gamma_S}= s_{\Gamma_S}
\label{fig:gradient-neumann}
 \end{equation}
 Both these Dirichlet and Neumann boundary conditions have been validated in previous works \cite{Ghigo,Limare}, showing at least second-order convergence for the multigrid Poisson-Helmholtz solver in presence of embedded boundaries.

\subsubsection{Navier boundary condition}

As for the Dirichlet condition, taking into account the Navier boundary condition Eq.~\eqref{eq:Navier}, is not straightforward since we need to express the embedded gradient on a fixed Cartesian grid while taking into account the local Navier boundary condition.
Considering the tangential and normal local coordinates $(\bar{\tau}, \nn)$ for a 2D system, attached to the solid rigid boundary, we write:
\begin{equation}
\grad \vv\cdot\nn= 
 \left(
\begin{array}{c}
\frac{\partial u_{\tau}}{\partial n} \\
\frac{\partial u_{n}}{\partial n}
\end{array}
\right)
\label{eq:local-normal}
\end{equation}
The transformation between the local system ($\bar{\tau}, \nn$) and the fixed Cartesian system ($\bar{e_x},\bar{e_y}$) is simply given by:
\begin{equation}
\grad \vv\cdot\nn^{x,y} = R^{\theta}\grad \vv\cdot\nn^{\bar{\tau}, \nn}
\label{eq:rotation-theta}
\end{equation}
 with
\begin{equation}
R^{\theta}= \left(
\begin{array}{cc}
  \cos\theta&  -\sin\theta \\
  \sin\theta&  \cos\theta
\end{array}
\right)
\text{or simply}
\left(
\begin{array}{cc}
  \tau_x &  -\tau_y\\
 \tau_y &  \tau_x
\end{array}
\right)
\end{equation}
We can then express the embedded gradient on $\Gamma_S$ for each velocity component:
\begin{subequations}
\begin{equation}
{\grad \vvx|_{\Gamma_S}} =  \tau_x \frac{\partial\utau} {\partial n} - \tau_y \frac{\partial \un}{\partial n}
\label{eq:gradux}
\end{equation}
\begin{equation}
{\grad \vvy|_{\Gamma_S}} =  \tau_y \frac{\partial\utau} {\partial n} + \tau_x\frac{\partial \un}{\partial n}
\label{eq:graduy}
\end{equation}
\label{eq:Navier-component}
\end{subequations}
Following the Navier boundary condition in Eq.~\eqref{eq:Navier}, and considering  $\bar{U}_{s,\tau}=0$ (motionless embedded boundary), $\displaystyle{\frac{\partial\utau} {\partial n}}$ and $\displaystyle{\frac{\partial\un} {\partial n}}$ are \textit{a priori} known at the interface $\Gamma_S$ since:
\begin{equation}
\left\{
\begin{array}{cc}
\frac{\partial u_{\tau}}{\partial n}= \frac{u_{\tau}}{\lambda}\\
u_n = 0
\end{array}
\right.
\label{eq:Navier-condition}
\end{equation}
Given the relation $u_n = 0$, we use the second-order gradient calculation Dirichlet method \eqref{eq:dirichlet-gradient} to evaluate $\displaystyle{\frac{\partial\un} {\partial n}}$.
The term $\displaystyle{\frac{\partial\utau} {\partial n}}$ can be approximated using the same principle,
\begin{equation}
\frac{\partial\utau} {\partial n} \approx \frac{1}{d_2-d_1}\left[\frac{d_2}{d_1}(\utau_{\Gamma}-{\utau}^{I_1})-\frac{d_1}{d_2}(\utau_{\Gamma}-{\utau}^{I_2})\right]
\label{eq:utau-dirichlet}
\end{equation}
But here, the value of the tangential velocity $\utau_{\Gamma}$ needs to be determined. Following
 \cite{JOHANSEN1998} we can write the relation \eqref{eq:utau-dirichlet}
\begin{equation}
\frac{\partial\utau} {\partial n} = a_0 \utau_{\Gamma} + a_1\utau^{I_1} + a_2  \utau^{I_2}
\label{eq:relation-colella}
\end{equation}
where the coefficients $a_0$, $a_1$ and $a_2$ are directly determined by \eqref{eq:utau-dirichlet}. Combining equations \eqref{eq:Navier-condition}- \eqref{eq:relation-colella} , we can eliminate $\frac{\partial\utau} {\partial n}$ and deduce the following relation to determine $\utau_{\Gamma}$
\begin{equation}
\utau_{\Gamma} = \frac{\lambda (a_1\utau^{I_1} + a_2  \utau^{I_2})}{1 - \lambda a_0}.
\label{eq:utau-dirichlet-bis}
\end{equation}
Given the value of $\utau_{\Gamma}$, the Dirichlet gradient calculation Eq.~\eqref{eq:dirichlet-gradient} is used again to calculate $\displaystyle{\frac{\partial\utau} {\partial n}}$.
Note that this method degenerates naturally into a Dirichlet boundary condition when $\lambda \rightarrow 0$ and into a free-slip condition when $\lambda \rightarrow \infty$, so that is can be used formally for any value of $\lambda$.
The generalization to 3D is also straightforward.

\subsubsection{Validation of the Navier boundary condition implementation}
\begin{figure}[!h]
 \centering
  \includegraphics[trim={3.5cm 5cm 2cm 2cm},clip,width=0.7\linewidth]{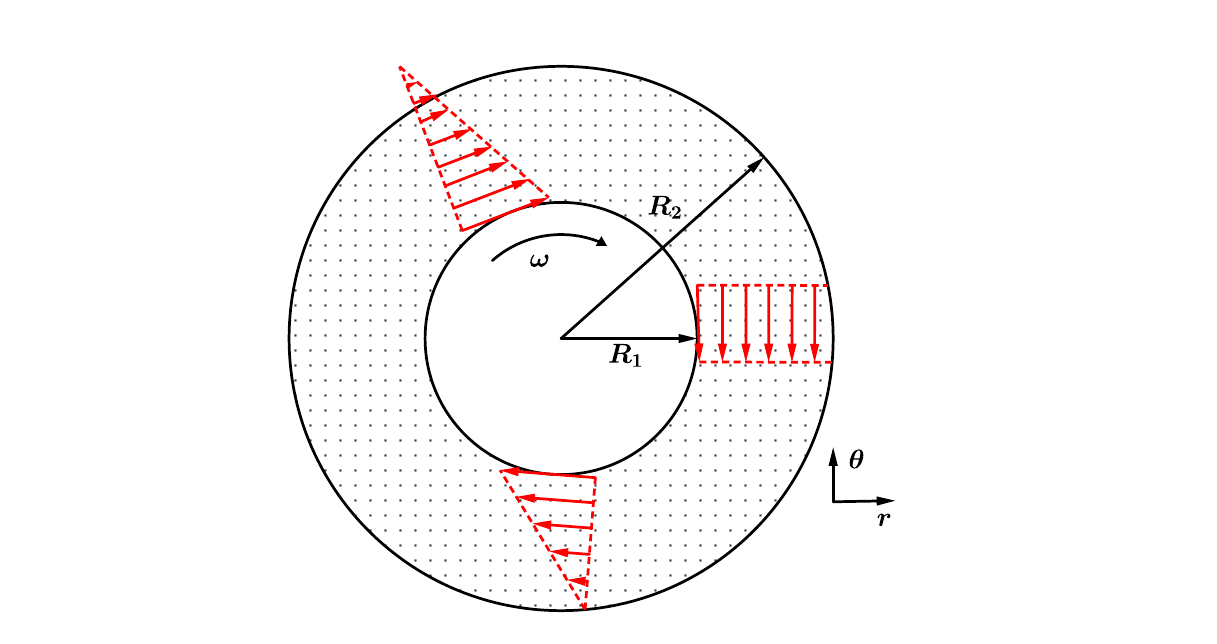}
  \caption{Schematic representation of the Taylor-Couette flow with Navier, free slip or Dirichlet boundary condition imposed at the outer fixed cylinder.}
  \label{fig:taylor-couette}
\end{figure}

To validate our method, we solve the (Stokes) Couette flow between two rotating cylinders using the Dirichlet, Navier or free slip boundary condition with embedded solid boundaries for a single phase flow.
We set the geometry of the embedded boundary as two cylinders of radii $R_2 = 0.5$ and $R_1 = 0.25$ (see fig.~\ref{fig:taylor-couette}).
Here, the outer cylinder is fixed and the inner cylinder is rotating with an angular velocity $\omega$.
For the free-slip boundary condition (outer cylinder free of viscous stresses), the analytical angular velocity $u_\theta$ is given by:
\begin{equation}
u_\theta(r) = \frac{\omega {R_1}^2 r}{{R_2}^2 + {R_1}^2}\left(\frac{{R_2}^2}{r^2} + 1 \right).
\label{eq:u_thetaslip}
\end{equation}
For a Navier boundary condition on the outer cylinder, we have:
\begin{equation}
u_\theta(r) = \frac{\omega {R_1}^2 }{ {R_1}^2(R_2 + \lambda) + {R_2}^2(\lambda - R_2)  }\left( (R_2 + \lambda) r + \frac{{R_2}^2 (\lambda - R_2)}{r} \right).
\label{eq:u_thetanavier}
\end{equation}
The simulation is performed in a two-dimensional box of size $D=2R_2$. Adaptive mesh refinement is not activated and we consider only a uniform Cartesian grid with respectively $N=16, 32, 64$ and $128$ grid points within the box size.
\begin{figure}[!h]
\begin{subfigure}[t]{0.49\textwidth}
  \centering
  \includegraphics[width=\textwidth]{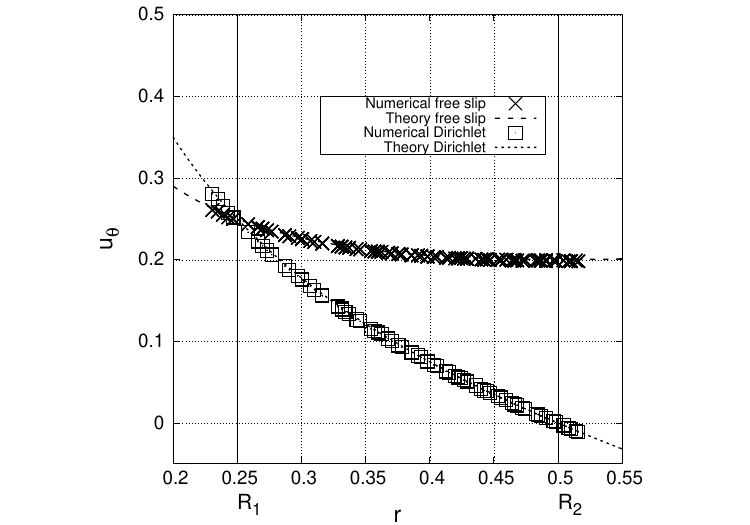}
  \caption{}
    \label{fig:free-slip}
  \end{subfigure}
  \begin{subfigure}[t]{0.49\textwidth}
  \centering
  \includegraphics[width=\textwidth]{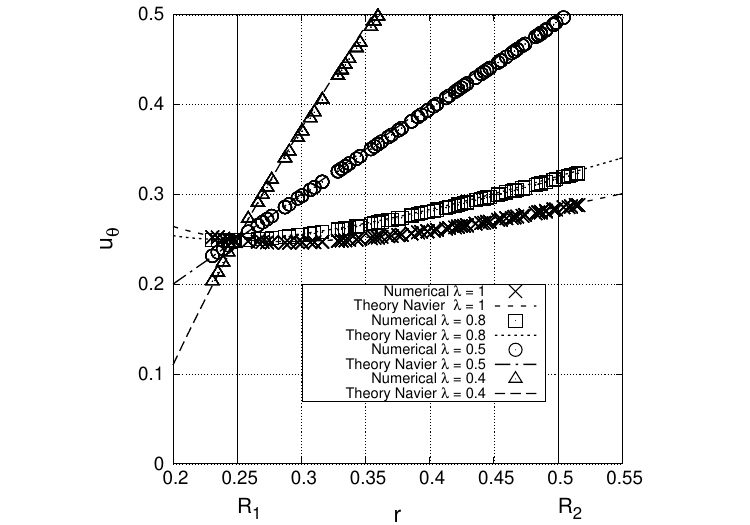}
  \caption{}
  \label{fig:Navier-slip}
  \end{subfigure}
  \caption{Comparison between the numerical and analytical angular velocity $u_\theta$ as a function of the radius $r$, $N=32$, (a) Free-slip and no-slip cases, (b) Navier slip case with different values of $\lambda = 0.4, 0.5, 0.8$ and $1$. See also \url{http://basilisk.fr/sandbox/tavares/test_cases/slip.c}}
\end{figure}
Figure \ref{fig:free-slip} illustrates the comparison between the numerical solution for $N=32$ and the analytical solution of equation \eqref{eq:u_thetaslip} in the free slip case. We observe very good agreement between both numerical and analytical solutions. The same comparison is carried out for the Navier boundary condition considering different values of $\lambda$, respectively $ 0.4, 0.5, 0.8$ and $1$. The agreement between the numerical and the theoretical solutions is also good whatever the value $\lambda$.
To further quantify the differences between numerical and analytical solutions, we define the errors:
\begin{equation}
E_1=  \frac{1}{N}\sum_{i=1}^{N}\left|({u_\theta^{ref}} - u_\theta^{num})\right|
\label{eq:erreur_1}
\end{equation}
and
\begin{equation}
E_\infty=  \max\left|({u_\theta^{ref}} - u_\theta^{num})\right|
\label{eq:erreur_infty}
\end{equation}
where ${u_\theta^{ref}}$ is the analytical velocity Eq.~\eqref{eq:u_thetaslip}-\eqref{eq:u_thetanavier} and $u_\theta^{num}$ the numerical velocity.

\begin{table}[]
\centering
\begin{adjustbox}{max width=\textwidth}
\begin{tabular}{ccccc|cccc}
\hline
  & \multicolumn{4}{c}{ Free slip} & \multicolumn{4}{c}{Dirichlet}               \\
 N   & $E_1$    & Order &    $E_\infty$      & Order & $E_1$        & Order & $E_\infty$         & Order \\
\hline \hline
16  & 6.08E-03 & -     & 1.33E-02 & -  & 4.54E-03  & -  & 2.28E-02  & -
 \\
32  & 1.36E-03 & 2.160 & 2.70E-03 & 2.300 & 1.07E-03 & 2.085 & 4.52E-03 & 2.335 \\
64  & 2.79E-04 & 2.285 & 7.53E-04 & 1.842 & 2.61E-04 & 2.035 & 1.25E-03 & 1.854 \\
128 & 6.77E-05 & 2.043 & 1.72E-04 & 2.130 & 5.98E-05 & 2.126 & 2.90E-04 & 2.108 \\
256 & 1.54E-05 & 2.136 & 5.10E-05 & 1.754 & 1.52E-05 & 1.976 & 8.07E-05 &1.845 \\
\hline
\end{tabular}
\end{adjustbox}
\caption{Dirichlet and free slip cases: Error $E_1$ Eq.~\eqref{eq:erreur_1}, $E_\infty$ Eq.~\eqref{eq:erreur_infty}  and order of convergence.}
\label{tab:slip-case}
\end{table}
Considering the free slip case, we can notice in table \ref{tab:slip-case} that the numerical solution converges to the theory at 2nd order for $E_1$ and $E_\infty$.
For the Navier slip case (table \ref{tab:Navier-slip-case}), the numerical solution also converges to the theoretical values but is sensitive to the value of $\lambda$. As the value of $\lambda$ is decreased (from 1 to 0.5), the order of convergence slowly drops from $2$ to $\approx 1.4$ on average for $E_1$. The same trend applies for $E_\infty$ in this case.
\begin{table}[]
\centering
\begin{adjustbox}{max width=\textwidth}
\begin{tabular}{ccccc|cccc}
\hline
  & \multicolumn{4}{c}{ $\lambda = 0.5$} & \multicolumn{4}{c}{ $\lambda = 1$}               \\
 N   & $E_1$    & Order &    $E_\infty$      & Order & $E_1$        & Order & $E_\infty$         & Order \\
\hline \hline
16 & 1.53E-03 & - & 2.87E-03 & - & 2.87E-03 & - & 2.87E-03 & -  \\ 
32 & 4.80E-04 & 1.672 & 1.17E-03 & 1.295 & 1.17E-03 & 1.295 & 1.17E-03 & 1.295  \\ 
64 & 1.00E-04 & 2.263 & 2.55E-04 & 2.198 & 2.55E-04 & 2.198 & 2.55E-04 & 2.198  \\ 
128 & 3.67E-05 & 1.446 & 1.26E-04 & 1.017 & 1.26E-04 & 1.017 & 1.26E-04 & 1.017  \\ 
256 & 3.70E-06 & 3.310 & 3.36E-05 & 1.907 & 3.36E-05 & 1.907 & 3.36E-05 & 1.907  \\ 
\hline
\end{tabular}
\end{adjustbox}
\caption{Navier slip case with $\lambda=0.5$ and $1$: Error $E_1$, $E_\infty$ and order of convergence.}
\label{tab:Navier-slip-case}
\end{table}

\subsection{Hybrid VOF/embedded boundary method}\label{sec1:subsubsec3}
We consider now a three phases system with a liquid-gas interface namely the liquid ($\Omega_L$), the gas ($\Omega_G$) and the solid ($\Omega_S$) (see Fig.~\ref{fig:vof-embed-exact} illustrating the typical configurations encountered).
\begin{figure}[!h]
\begin{subfigure}[t]{0.49\textwidth}
\includegraphics[trim={2cm 3cm 2cm 2cm},clip,width=1.01\linewidth]{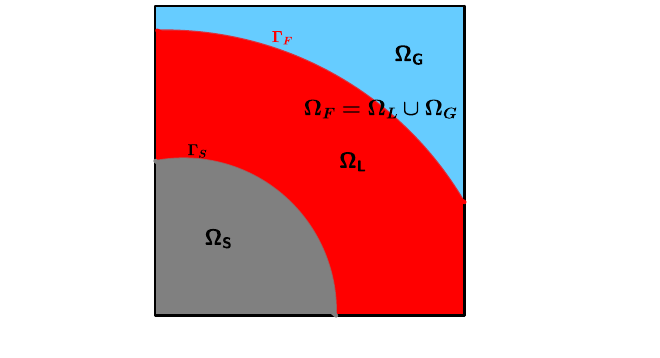}
\vspace{-0.5cm}
\caption{}
\label{fig:vof-embed-exact}
\end{subfigure}
\begin{subfigure}[t]{0.49\textwidth}
\includegraphics[trim={2cm 2.4cm 2cm 1.6cm},clip,width=1\linewidth]{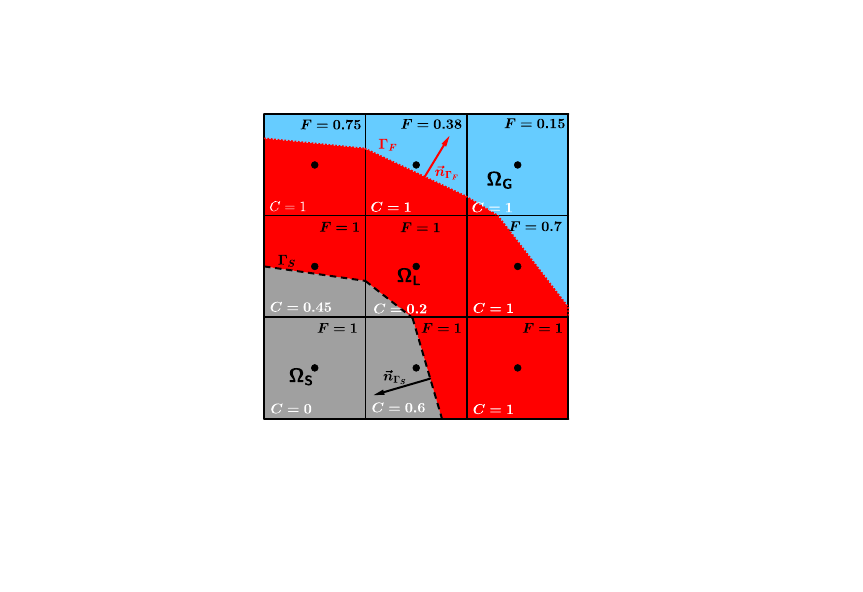}
\caption{}
\label{fig:vof-embed-config}
\end{subfigure}
\begin{subfigure}[t]{0.49\textwidth}
\includegraphics[trim={2cm 3cm 2cm 2cm},clip,width=1.05\linewidth]{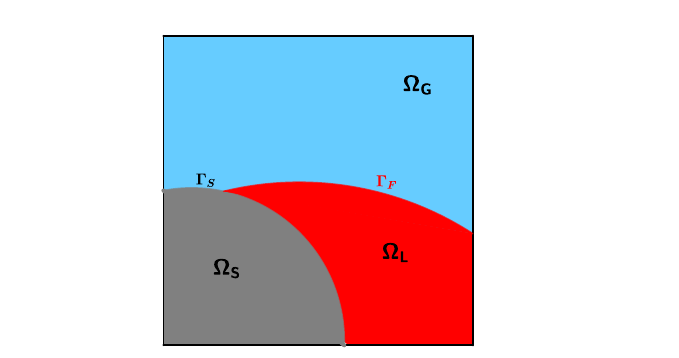}
\caption{}
\label{fig:vof-embed-exact}
\end{subfigure}
\begin{subfigure}[t]{0.49\textwidth}
\includegraphics[trim={2cm 2.7cm 2cm 1.8cm},clip,width=1\linewidth]{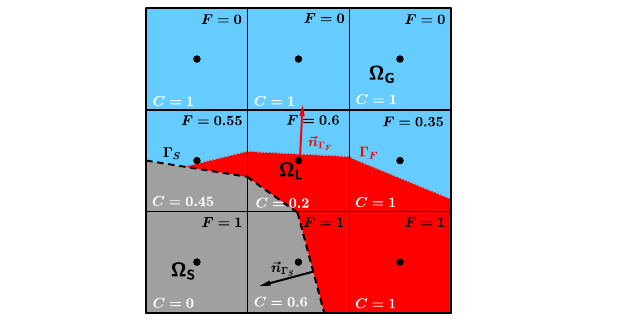}
\caption{}
\label{fig:vof-embed-config-bis}
\end{subfigure}
\caption{Rigid boundary ($\Omega_S$) embedded in a two-phase flow ($\Omega_L$), ($\Omega_G$): (a) Continuous solid, liquid and gas interfaces characteristics, (b) Volume fraction $F$ and $C$ in the discrete solid liquid-gas system (without intersection), (c) Continuous solid, liquid and gas interfaces with intersection, (d) discrete solid, liquid and gas system with intersection.}
\end{figure}

To solve such problems, we can {\it a priori} simply couple the two methods described in the previous section, the Volume of Fluid one for the liquid-gas, and the embedded boundary method for the solid-fluid system. 
As stated in (see fig.\ref{fig:vof-embed-exact},\ref{fig:vof-embed-config}) the flow must now account for two volume fractions: $F$ to track the liquid-gas interface and $C$ for embedded solid cells.
Both volume fractions $F$ and $C$ are defined in the whole domain $\Omega_G\cup\Omega_L\cup\Omega_S$ following {\it a priori}:
\begin{equation}
F(\xx,t) = \left\{
\begin{array}{cc}
0 &\mbox{in\, $\Omega_G$}\\
1 &\mbox{in\, $\Omega_L\cup\Omega_S$}
\end{array}
\right .
\label{eq:F}
\end{equation}
whereas
\begin{equation}
C(\xx,t) = \left\{
\begin{array}{cc}
0 &\mbox{in\, $\Omega_S$}\\
1 &\mbox{in\, $\Omega_L\cup\Omega_G$}
\end{array}
\right .
\label{eq:C}
\end{equation}
It is important to note that the value of liquid-gas function $F$ in the domain where $C=0$ can in fact take any value since it is not used so far in the equations.
Moreover, since the embedded boundary method only intervenes in the fluid cells near the cut-cells, the coupling is at work only in this region.
Therefore, in the presence of a solid boundary, the solution of the two-phase flow system (Eqs.~\eqref{eq:Navier-stokes},\eqref{eq:volume-fraction}) described in section \ref{sec:sec1} needs to be adapted only near the solid boundary, since it will not at all be considered in the cells where $C=0$.

Let us first consider consider the VOF advection, to be performed in the full fluid cells but also in the ``mixed'' cells ($0<F<1$ and $0<C<1$, $\Omega_G\cap\Omega_L\cap\Omega_S$): a simple approximation, which we use here, is to ignore the solid volume fraction.

\begin{figure}[!h]
\begin{subfigure}[t]{0.49\textwidth}
\centering
\includegraphics[trim={6.5cm 6.5cm 4cm 4cm},clip,width=1.2\linewidth]{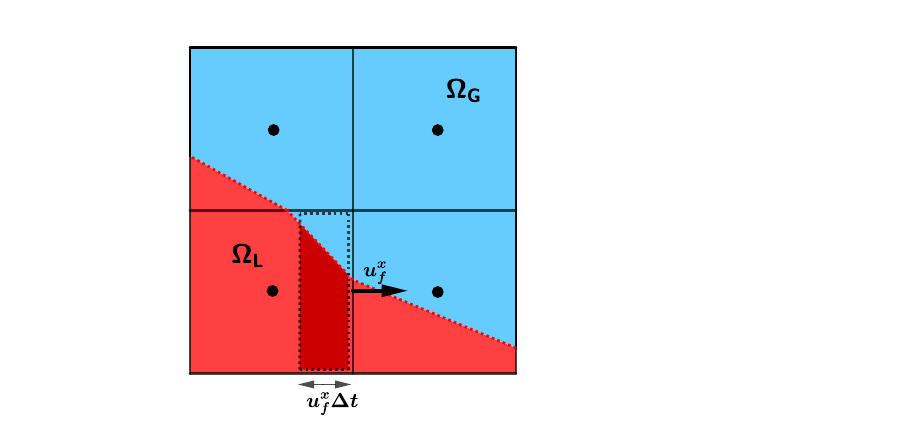}
\caption{} \label{fig:flux-fluid-cell}
\end{subfigure}
\begin{subfigure}[t]{0.49\textwidth}
\includegraphics[trim={6.5cm 6.5cm 4cm 4cm},clip,width=1.2\linewidth]{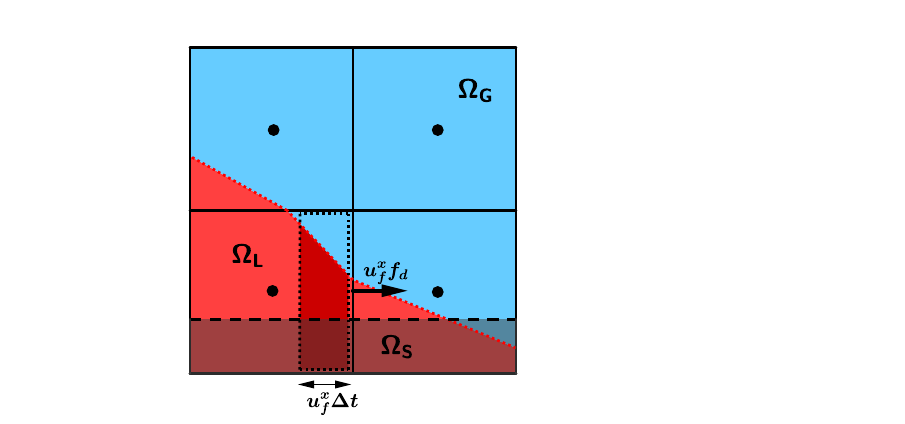}
\caption{}
\label{fig:flux-cut-cell}
\end{subfigure}
\caption{Geometric estimation of the flux $G$: (a) in a two-phase cell, (b) in a mixed (or three-phase) cell.}
\end{figure}

This ensures robustness of the VOF advection and preserves the conservation of the total fluid mass, if the latter is computed ignoring the volume occupied by the solid in ``mixed'' cells. The solid can then be interpreted as ``slightly porous in surface'' i.e. able to retain some of the fluid, but only within a thin shell (i.e. thinner than the mesh size). So we expect this method to be no more than first-order in convergence. It will be therefore useful to quantify this ``local mass absorption'' and this is done in the validation section \ref{sec:sec3} for different test cases. We observe that it can reach $5\%$ in the most severe configurations (smallest contact angles).
A rough way to compensate for this mass absorption could be to add or subtract the mean mass absorption from each cell at the fluid interface \cite{PATEL2017}. In that case, one must assume that the mass compensation does not affect the global flow, thus the local mass absorption must be small.
A higher-order approach could also be employed by modifying the flux in ``mixed'' cells as illustrated in Figure \ref{fig:flux-fluid-cell} to calculate the flux $G=F \vvx_f$, including the solid volume.
Figure \ref{fig:flux-cut-cell} shows the horizontal VOF flux calculation in a ``mixed'' cell. Here, the dotted line delimits the flux going to the right-hand neighbor. Even if the normal face velocity is weighted by the face fraction $\vvx_f f^d$, the total flux estimation includes the solid region.
A better estimate of this flux should take into account the solid region area in the calculation.
In the example given in Fig.~\ref{fig:flux-cut-cell}, the exact geometric flux estimation is straightforward as the solid region is a rectangle.
In more general cases including 3D configurations, this estimation is less trivial and requires a specific algorithm to account for the possible orientations of general solid geometries. This problem is not addressed here but would be an interesting improvement of the coupled VOF/embedded boundary method.

The other important issue when coupling VOF and embedded boundary is the boundary conditions to impose on the liquid-gas volume fraction $F$. For the single flow cut cells (where $F=0$ or $F=1$, with $0<C<1$), as illustrated on figures \ref{fig:vof-embed-exact} and \ref{fig:vof-embed-config}, no special action is required, beside taking into account the correct viscous boundary conditions explained above. On the other hand, the situation is more complicate for mixed cells since the reconstruction of the interface is crucial to estimate accurately the fluid fluxes and the surface forces. This corresponds also to the presence of a triple point/line in the cell as illustrated in Figure \ref{fig:vof-embed-config-bis}. In this configuration a contact angle boundary condition has to be explicitly set because of the non-conforming body description of the fluid and solid region. To do so, we use ``ghost cells'' taking advantage of the fact that the function $F$ can be arbitrarily set in the full solid cells (where $C=0$). 
The principle of the ghost-cell method is then to impose the values of $F$ in the full solid cells near the mixed cell in order to obtain the correct contact angle for the interface in the mixed cells.

\subsection{Contact angle boundary condition with the VOF/embedded boundary method}
In a three-phase system, the contact angle computation is one of the main challenges since it plays a major role on surface tension effects controlled by the motion of contact lines and therefore the wetting/dewetting of the solid.
Our objective is to propose a simple methodology which captures the physics of the wetting by imposing a prescribed contact angle. This angle could either be static $\theta_s$ or dynamical $\theta_d(t)$ with an instantaneous dynamical angle determined with an appropriate model \cite{SHIKHMURZAEV_1997,Legendre_2015,Afkhami_2009}.
In what follows, the algorithm to enforce this contact angle (noted $\theta_s$ for simplicity further on, with no loss of generality) is described.
The contact angle $\theta_s$ is thus imposed at the embedded boundary wall by setting the corresponding fluid volume fraction $F$ in $\Omega_S$, using the concept of ghost cells \cite{asghar2023,PATEL2017,O_Brien_2018}. 
This is done in several steps:
\begin{itemize}
\item First, according to a prescribed value of $\theta_s$, the properly-oriented normal $\nn_{\theta}$ at the liquid-gas-solid intersection is determined:
\begin{equation}
\nn_{\theta} = \nn_{\Gamma_S}\cos\theta_s + \tt_{\Gamma_S}\sin\theta_s
\label{eq:normal-contact}
\end{equation}
\begin{figure}[!h]
\begin{subfigure}[t]{0.49\textwidth}
    \centering
    \includegraphics[trim={4cm 7cm 0cm 4cm},clip,scale=0.6]{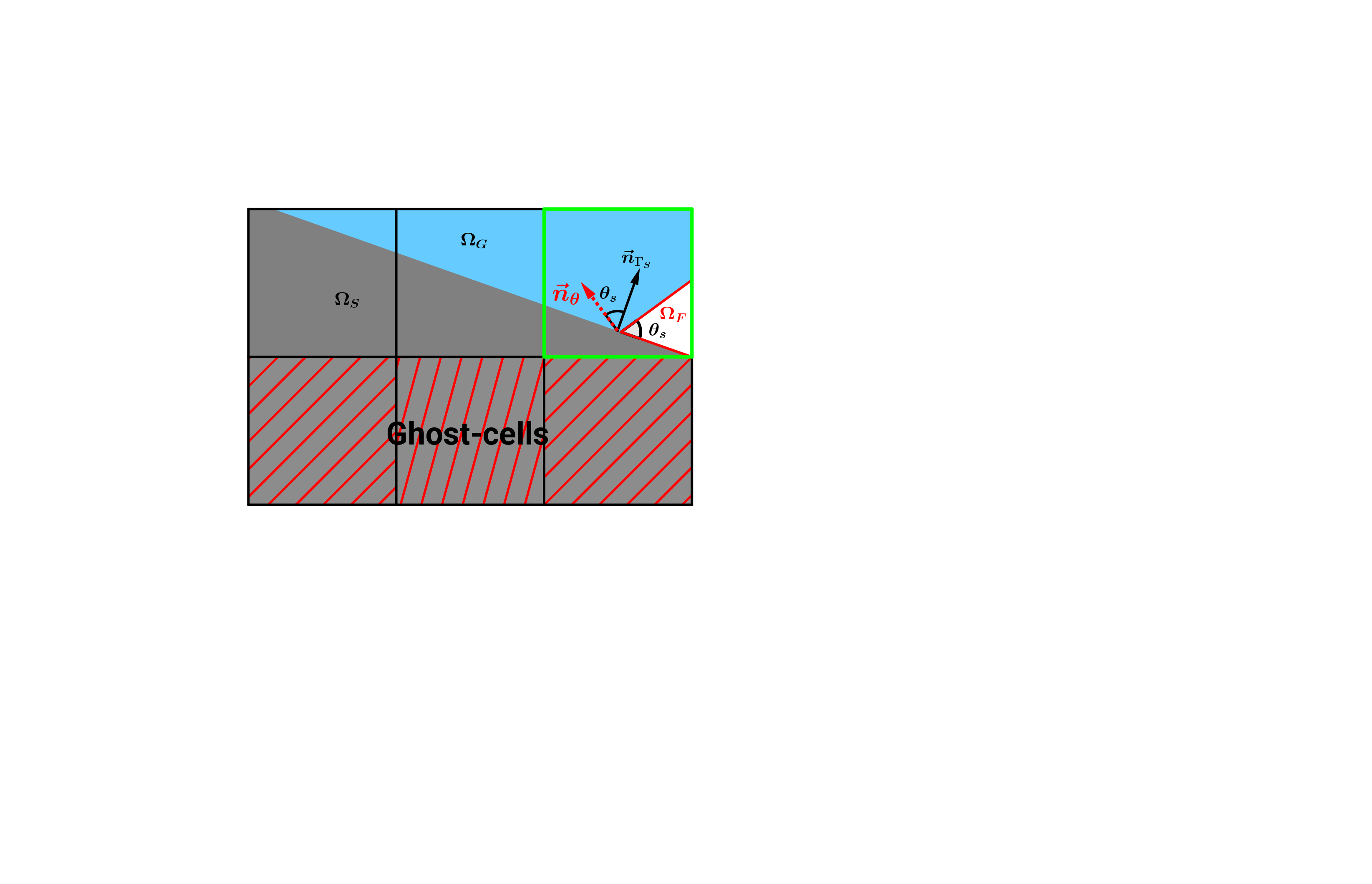}
    \caption{}
    \label{fig:scheme-algorithm}
\end{subfigure}
\begin{subfigure}[t]{0.49\textwidth}
    \centering
    \includegraphics[trim={4cm 7cm 0cm 4cm},clip,scale=0.6]{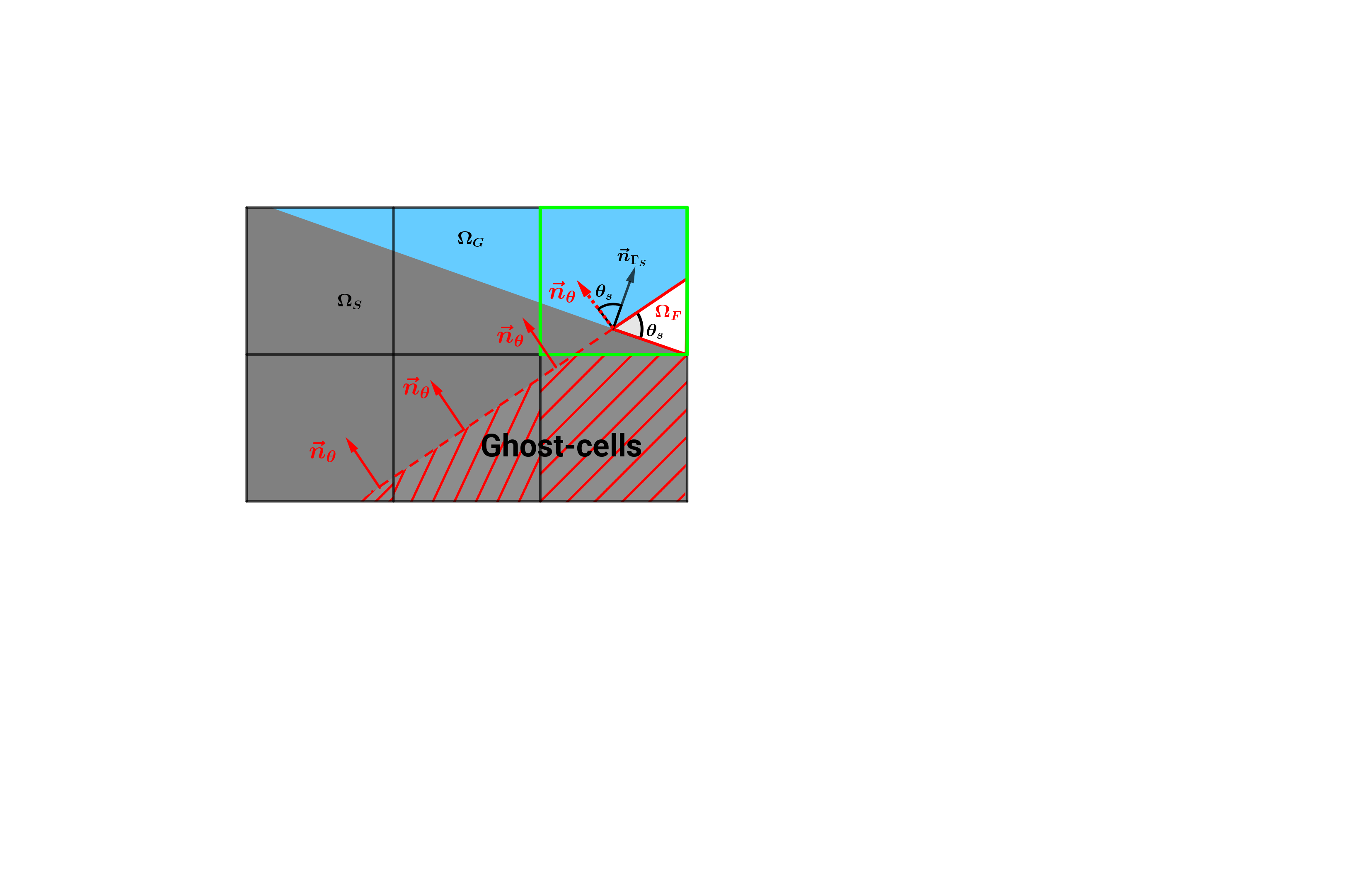}
    \caption{}
    \label{fig:scheme-algorithm1}
\end{subfigure}
\caption{Contact angle algorithm description: (a) The normal $\nn_{\theta}$ is reconstructed following Eq.~\eqref{eq:normal-contact} (b) The volume fraction in the ghost cells $F_{G}$ is calculated (Tab.~\ref{tab:algorithme 2}), and the boundary condition at the contact line for the prescribed value of $\theta_s$ is eventually imposed.}
\end{figure}
For ``mixed'' cells ($0<C<1$ and $0<F<1$), $\nn_{\theta}$ is calculated so that the intercept $\alpha_{\theta}=\func(F, \nn_{\theta})$ can be determined following the procedure described in section \ref{sec:sec2subsubsec1}.
\begin{table}[h!]
\centering
\begin{tabular}{l}
\hline
\textbf{Algorithm 1}\tabularnewline
\hline
\begin{minipage}{\linewidth}
\begin{lstlisting}[mathescape=true]
for all cells
 if ($0<C<1$ and $0<F<1$) then // potential triple phase cell
  $\nn_{\theta} = \nn_{\Gamma_S}\cos\theta_{\Gamma_S} + \tt_{\Gamma_S}\sin\theta_{\Gamma_S}$ // normal at the triple point
  $\alpha_{\theta} = \func(F, \nn_{\theta})$ // intercept at the triple point cell
 end if
end for
\end{lstlisting}
\end{minipage} \tabularnewline
\hline
\end{tabular}
\caption{Interface geometric characteristics calculation at the triple point.}
\label{tab:algorithme 1}%
\end{table}
The algorithm (see Tab.~\ref{tab:algorithme 1}) summarizes the first steps of the contact angle procedure with the calculation of the interface geometric properties $\nn_{\theta}$ and $\alpha_{\theta}$ of the cells with three phases.
\item We then consider the neighboring ghost cells ($C=0$) where the volume fraction $F$ has to be modified to account for the contact angle condition.
If the neighbor of a ghost-cell within two cells is a cell with three phases ($0<C<1$ and $0<F<1$), a coefficient to weight the fluid volume fraction in the ``ghost-cells" depending on how well the ``mixed'' cell is defined, is computed.
We assume that a ``mixed'' cell is better defined when $C\approx F\approx 0.5$.
Knowing the values of the intercept $\alpha_{\theta}$ and the normal $\nn_{\theta}$ in these cells (see Tab.~\ref{tab:algorithme 2}), the prescribed interface (accounting for the prescribed value of $\theta_s$) is extrapolated into the considered ``ghost cells'' and a new fluid volume fraction $(F=\func(\alpha_{\theta}, \nn_{\theta}))$ is computed.
The final volume fraction $F_G$ in the ``ghost cells'' is a weighted average of the volume fractions reconstructed from the cells neighboring a cell with three phases. 
\begin{table}[h!]
\centering
\begin{tabular}{l}
\hline
\textbf{Algorithm 2}\tabularnewline
\hline
\begin{minipage}{\linewidth}
\begin{lstlisting}[mathescape=true]
for all cells
 if (C == 0) then // in the ghost cells
  $w_{cell}=0$, $w_{total}=0$ // weight to estimate the volume fraction reconstruction
  for all neighbors within 2 cells
   if ($0<C<1$ and $0<F<1$) then // potential triple phase cell
    $w_{cell}=C\times (1. - C)\times F\times(1. - F)$
    $w_{total}=\sum w_{cell}$
    $F_{G} = w_{cell}\times f(\nn_{\Gamma_F}, \alpha_{\theta})$
   end if
  end for
  $F_{G} = F_{G}/w_{total}$
 end if
end for
\end{lstlisting}
\end{minipage} \tabularnewline
\hline
\end{tabular}
\caption{Fluid volume fraction calculation inside the ghost cells.}
\label{tab:algorithme 2}%
\end{table}
\end{itemize}
This contact angle procedure allows to apply the proper boundary condition at the liquid-gas-solid intersection by modifying only the values of the fluid volume fraction $F$ inside the ``ghost cells''. 
A shown in fig.~\ref{fig:scheme-algorithm1} a neighborhood of two ``ghost cells''around the triple point is required to obtain an accurate definition of the contact angle $\theta_s$. 

\subsubsection{Contact line dynamics}
The motion of the contact line along a no-slip solid surface involves a mathematical paradox.  
Indeed, the no-slip boundary condition is known to introduce a ``stress singularity" at the contact line since the stress generated by the moving contact line diverges with grid refinement. Several authors in the literature have proposed various models to deal with this singularity \cite{Afkhami_2009,Fullana_2020,Afkhami_2022}. 
Using the VOF method intrinsically removes this contradiction as 
the volume fraction advection uses the face centered velocity which is not strictly zero, even if a zero velocity is imposed at the solid boundary. Thus, the contact line can move due the so-called ``numerical slip" \cite{Renardy_2001}.
This ``numerical slip" is linked to the grid size and tends toward the no-slip limit as the size of the mesh is decreased.
A way to control this mesh dependency while relaxing the stress singularity at the contact line is to use slip models. The most common slip model in the literature is the Navier slip law Eq:~\eqref{eq:Navier}.
In what follows, We will use either Dirichlet \eqref{eq:dirichlet-gradient} or Navier Eq:~\eqref{eq:Navier-component} boundary conditions coupled with the contact angle calculation (thus taking a fixed $\theta_s$) described above for the contact line dynamics.



\section{Validations tests}\label{sec:sec3}
In this section, we investigate different configurations in order to test and validate the implementation of the hybrid VOF/embedded boundary method. These test cases are designed to challenge the VOF/embedded coupling and particularly the boundary condition at the triple point/line with the geometrical approach detailed in the previous section.

The constant static contact angle $\theta_s$ is imposed and we start in general with an initial condition where the contact angle is $\theta_i \ne \theta_s$ so that we investigate the relaxation towards an equilibrium static configuration. Indeed, since the initial contact angle is out of equilibrium, the VOF/embedded dynamics should converge towards a situation where the contact angle is the static one. In most cases, analytical static configurations exist that we will compare to the numerical solution. We will investigate both 2D (plane) and 3D geometries.

We first simply study the ``numerical dynamics" that is using no-slip boundary conditions on the solid boundaries and no dynamical contact angle modeling. These conditions are known to mimic a (numerical) Navier slip boundary conditions at the solid boundaries~\cite{Afkhami_2009,DUPONT2010} and  we will investigate a classical dynamical model for the contact line dynamics using a slip length~\cite{Eggers2004}.
We will therefore consider droplets on various solid geometries: a cylinder, an horizontal plane and an inclined plane with or without gravity.
Except when explicitly stated, the density and viscosity ratios for the liquid-gas interface are set to $\rho_l/\rho_g=1$ and $\mu_l/\mu_g=1$ and the surface tension $\sigma=1$.
To ensure the stability of the simulations, the time step is constrained by the oscillation period of the smallest capillary wave given by $\Delta t = \sqrt\frac{(\rho_g+\rho_l)\Delta^3}{2\pi\sigma}$. Moreover, the maximum CFL number is set to $0.5$ for all cases.

\subsection{Droplet on an embedded cylinder}\label{subsec:subsec1}
\begin{figure}[!h]
\centering
\includegraphics[trim={2cm 3.5cm 2cm 4cm},clip,width=\linewidth]{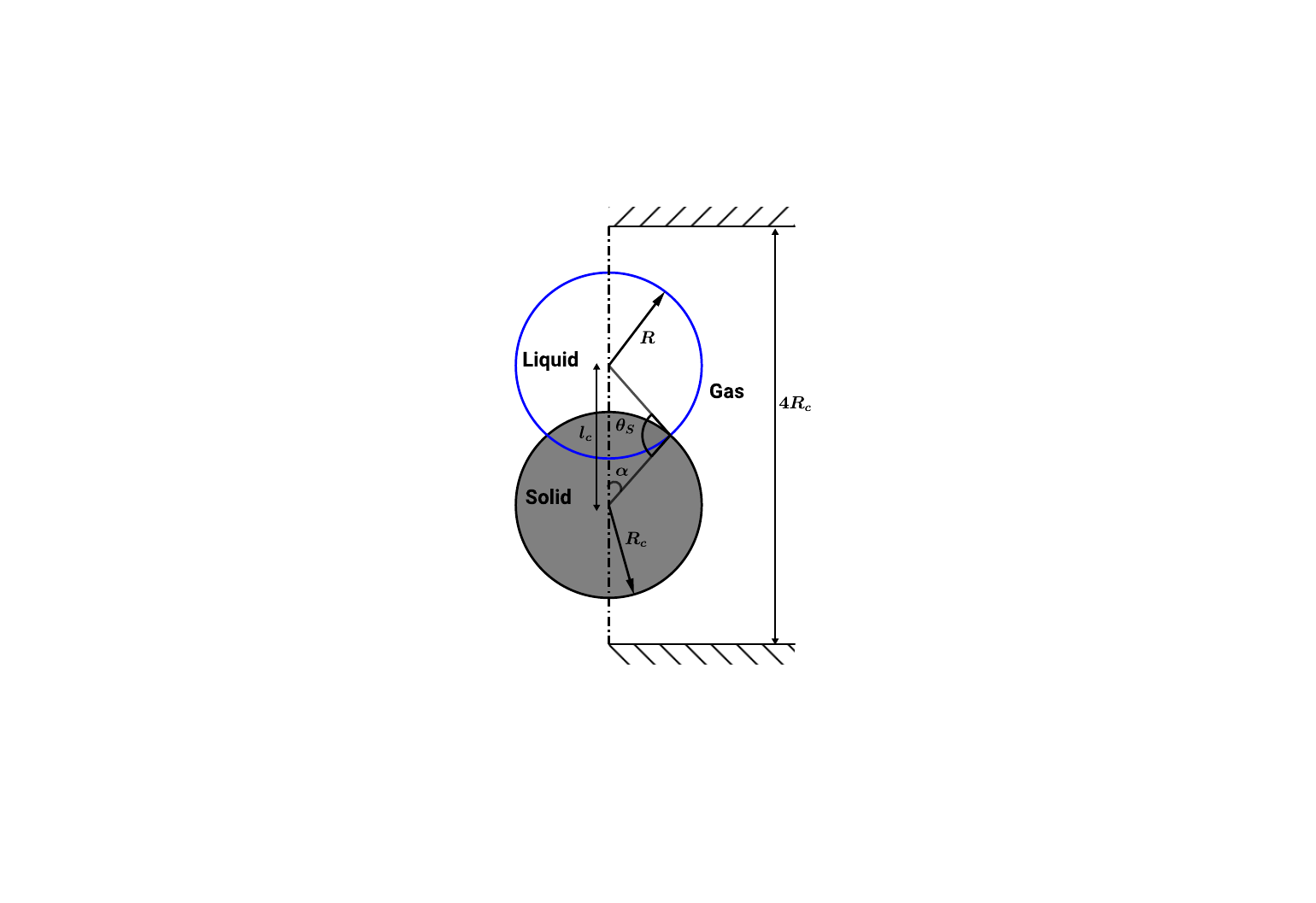}
\caption{Schematic representation of the initial configuration of the droplet on a cylinder.}
\label{fig:droplet-cylinder-init}
\end{figure}

\begin{figure}[!h]
\begin{subfigure}[t]{0.49\textwidth}
  \centering
  \includegraphics[width=\textwidth]{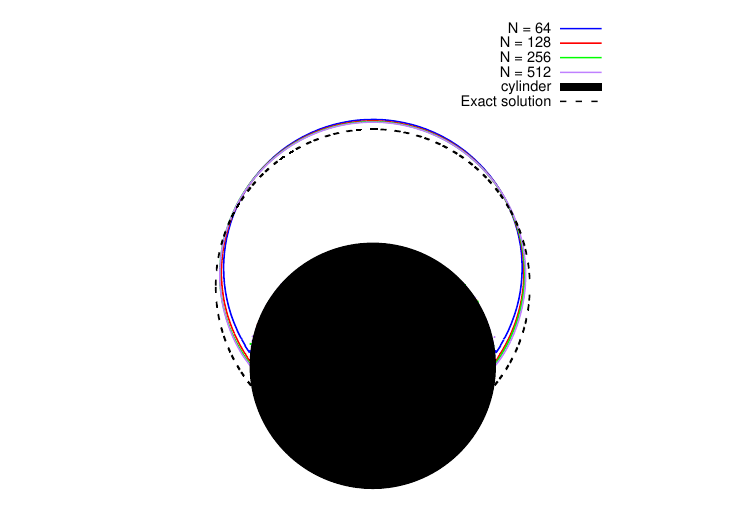}
  \caption{}
  \label{fig:cylinder-shape-convergence30}
  \end{subfigure}
  \begin{subfigure}[t]{0.49\textwidth}
  \centering
  \includegraphics[width=\textwidth]{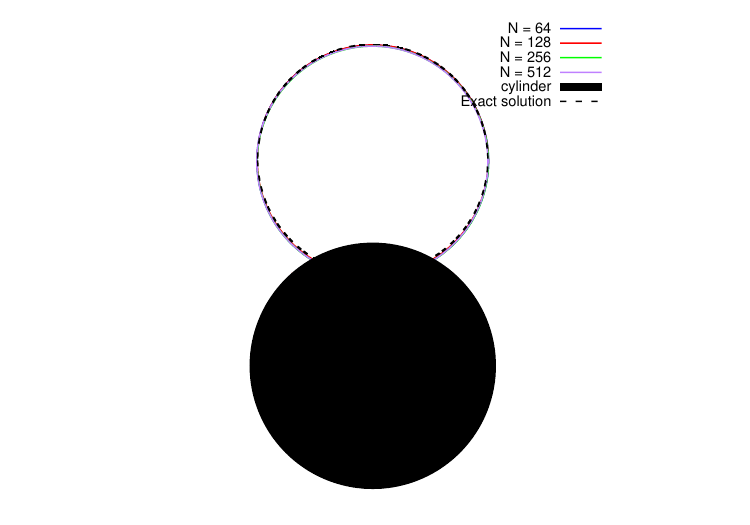}
  \caption{}
  \label{fig:cylinder-shape-convergence120}
  \end{subfigure}
  \caption{Convergence of the equilibrium droplet shape with respect to the grid refinement for $\theta_s = 30^\circ$ and $120^\circ$}
\end{figure}

We investigate in this test case the ability of the coupled VOF/embedded boundary method described above to impose a contact angle on a curved solid geometry.
The equilibrium shape of a droplet on a solid circular cylinder of radius $R_c=0.5$ is investigated here for different static contact angles $\theta_s$. As shown in Fig.~\ref{fig:droplet-cylinder-init}, a droplet of initial radius $R_0=0.5$ wets a cylinder with an initial contact angle $\theta_i = 90^\circ$.
In the absence of gravity, the equilibrium shape of the droplet can be obtained analytically.
For a two-dimensional droplet, the area corresponding to the equilibrium state is obtained by carving the cylinder out of a circle so that, the initial area $S_0$ of the droplet is given by:
\begin{multline}
S_0 = \pi R_0^2 - {R_c}^2\arccos\left(\frac{R^2 + {R_c}^2 -{l_c}^2}{2R l_c}\right) - {R_0}^2\arccos\left(\frac{R^2 + {l_c}^2 -{R_c}^2}{2R_0 l_c}\right) + \\\sqrt{\frac{(R_c + R -l_c)(R_c - R + l_c)(R_c + R + l_c)(-R_c + R + l_c)}{4}}
\end{multline}
The final radius of the droplet $R$ and the distance between the center of the cylinder and the droplet $l_c$ can be computed by finding the roots of the equation
\begin{equation}
f(l_c) = S_0 - (\alpha + \theta)R^2 + \alpha R_c^2 - RR_c\sin\theta
\label{eq:cylinder-surface}
\end{equation}
with
\begin{equation}
\alpha(l_c) = \arccos\left(\frac{R^2 + {R_c}^2 -{l_c}^2}{2R l_c}\right)
\label{eq:cylinder-angle}
\end{equation}
and
\begin{equation}
R(l_c) = R_c\cos\theta + \sqrt{{R_c}^2\cos\theta^2 - {R_c}^2 + {l_c}^2}
\label{eq:cylinder-radius}
\end{equation}
The simulation is performed in a two-dimensional box of size $2D\times2D$ with $D=2R_0$. It is enclosed by solid walls for all boundaries except the cylinder which is described using the embedded boundary method. The contact angle boundary condition is applied using the coupled VOF/embedded boundary method as described above. The adaptive mesh refinement is not enabled here so that we always have a Cartesian uniform grid with respectively $N=32, 64, 128$ and $256$ cells and $8, 16, 32$ or $64$ cells describing the initial radius $R_0$ of the drop (for instance $R=16\Delta$ for $N=64$).
Five different contact angles $\theta_s$, $30^\circ, 60^\circ, 90^\circ, 120^\circ$ and $150^\circ$ have been studied.
\begin{figure}[!h]
  \begin{subfigure}[t]{0.3\linewidth}
  \centering\includegraphics[width=2\linewidth]{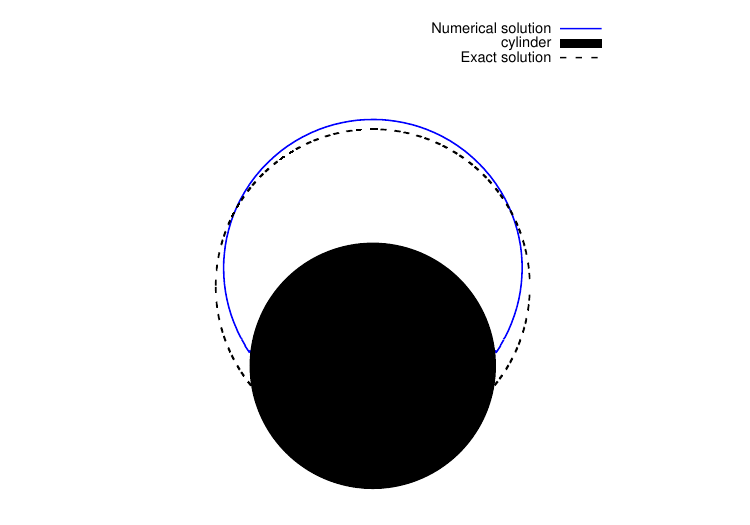}
  \caption{$\theta_s=30^\circ$}
  \end{subfigure}
  \begin{subfigure}[t]{0.3\linewidth}
  \centering\includegraphics[width=2\linewidth]{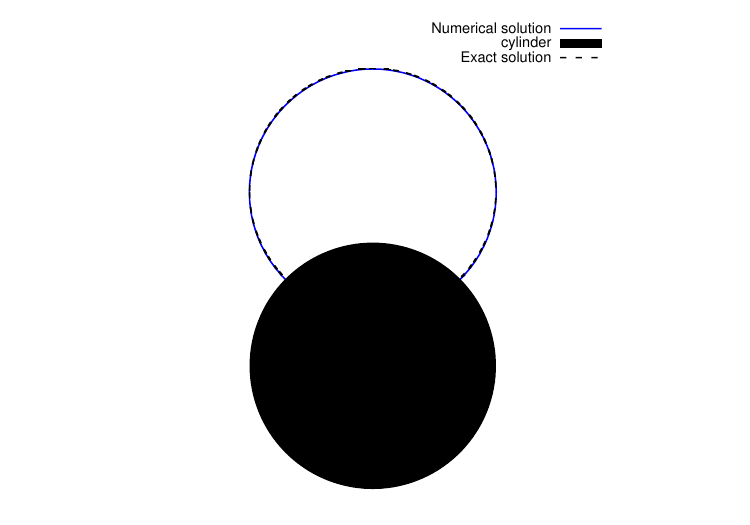}
  \caption{$\theta_s=90^\circ$}
  \end{subfigure}
  \begin{subfigure}[t]{0.3\linewidth}
  \centering\includegraphics[width=2\linewidth]{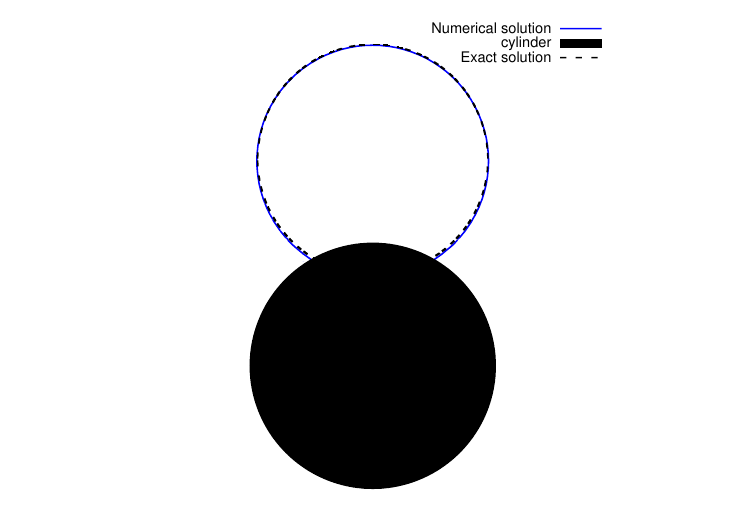}
  \caption{$\theta_s=120^\circ$}
  \end{subfigure}
  \caption{Comparison of the equilibrium droplet shapes against the analytical results for $30^\circ<\theta_s<150^\circ$, $N=64$. See also \url{http://basilisk.fr/sandbox/tavares/test_cases/droplet-cylinder-embed.c}.}
  \label{fig:cylinder-shape}
\end{figure}
The comparison between the numerical and the analytical equilibrium droplet shapes shows satisfactory qualitative agreement with the analytical solutions for most of the contact angles studied, as shown in Figures \ref{fig:cylinder-shape}. More precisely, the agreement is better for hydrophobic contact angles while it is less good for hydrophilic cases (see for instance $\theta_s= 30^\circ$). Figures ~\ref{fig:cylinder-shape-convergence30}-\ref{fig:cylinder-shape-convergence120}
shows the analytical and numerical solutions as the grid is refined, for $\theta_s= 30^\circ$ and $\theta_s= 120^\circ$.

For $\theta_s= 30^\circ$, the discrepancy between analytical and numerical solutions is clearer. However, it is difficult to state if the numerical solution slowly converges to the analytical one with grid refinement or not.
The lack of convergence is related to a numerical pinning of the contact line: depending on the mesh orientation with the embedded boundary, a situation where the VOF flux advection reduces to zero can occur and the contact line cannot relax anymore to the prescribed angle. This pinning effect is clearer for the test case of the droplet on an inclined plane (see sec.~\ref{subsec:subsec2}). This ``pinning effect'' can also be interpreted through the dependency of the ``numerical Navier slip'' condition on the details of the cut-cell configuration: for (nearly) full fluid cells close to the embedded boundary, the numerical Navier slip length is of the order of $\Delta/2$ while for (nearly) full solid cells this numerical slip length tends toward zero (i.e. no-slip).

\begin{figure}[!h]
  \centering
  \begin{subfigure}[t]{0.45\textwidth}
  \includegraphics[width=1.2\textwidth]{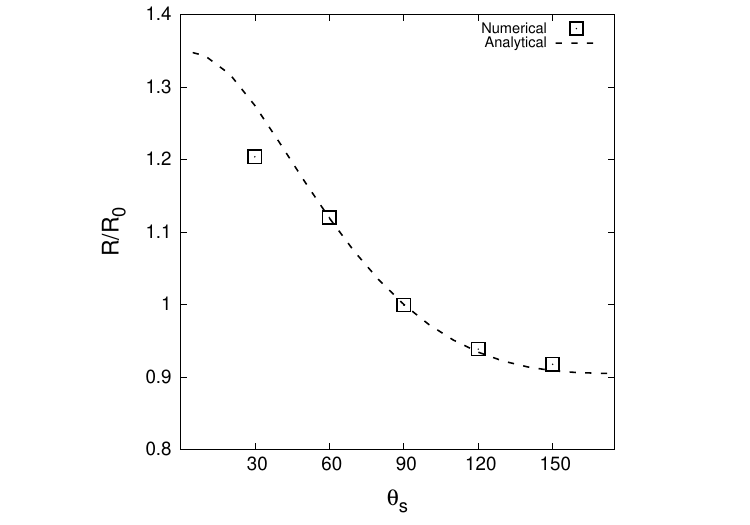}
  \caption{}
    \label{fig:cylinder-angle-shape}
    \end{subfigure}
    \begin{subfigure}[t]{0.45\textwidth}
    \includegraphics[width=1.2\textwidth]{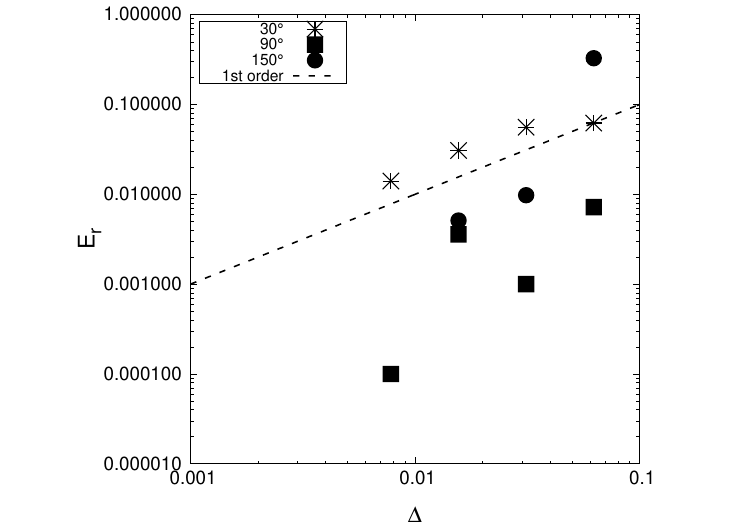}
    \caption{}
    \label{fig:cylinder-radius-convergence}
    \end{subfigure}
    \caption{(a) Dimensionless droplet radius $R/R_0$ evolution for the equilibrium shape of a droplet on a cylinder for $N=64$, (b) Relative error on the final droplet radius for $\theta_s=30^\circ$, $90^\circ$, $150^\circ$.}
\end{figure}

\begin{figure}[!h]
  \centering
 \begin{subfigure}[t]{0.45\textwidth}
  \includegraphics[width=1.2\textwidth]{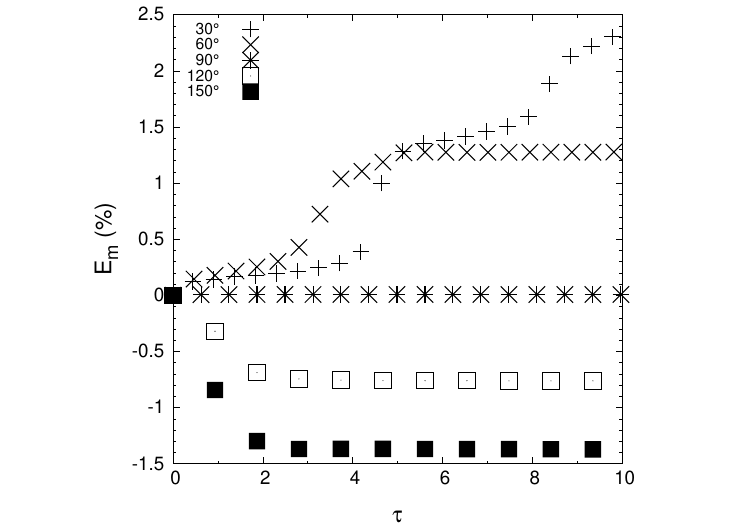}
  \caption{}
    \label{fig:cylinder-mass-variation}
    \end{subfigure}
    \begin{subfigure}[t]{0.45\textwidth}
    \includegraphics[width=1.2\textwidth]{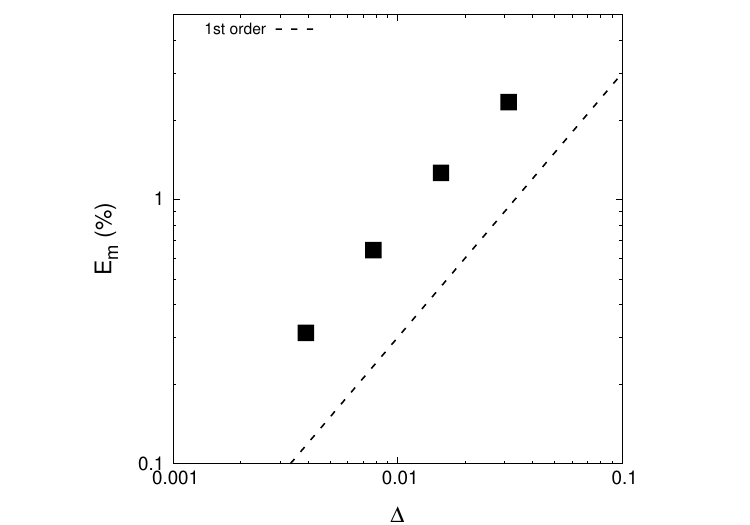}
    \caption{}
 \label{fig:cylinder-mass-convergence}
    \end{subfigure}
  \caption{(a) Relative mass absorption as a function of time and for different contact angles, for $N=64$. For each angle it converges at large times to an asymptotic value (not shown here for $\theta_s=30^\circ$); (b) Asymptotic relative mass absorption for $\theta_s=30^\circ$ with respect to the grid space $\Delta$.}

\end{figure}

To quantify the accuracy of our method, we also computed the numerical final radius of the droplet shown on figures \ref{fig:cylinder-angle-shape}.
Except for the smallest value of $\theta_s=30^\circ$, the numerical and analytical radius are in good agreement. To quantify it more precisely, we show on figure \ref{fig:cylinder-radius-convergence} the relative error on the radius as a function of the mesh size $\Delta$ for $\theta_s=30^\circ$, $90^\circ$ and $150^\circ$, exhibiting roughly the expected first-order convergence.
Furthermore, we also compute the relative mass variation during the simulation, plotted on figure \ref{fig:cylinder-mass-variation} for different contact angles $\theta_s$ for $N=64$. We observe that this mass variation converges to an asymptotic value at large times, when the droplet has reached its equilibrium state.

Figure \ref{fig:cylinder-mass-convergence} shows this relative mass absorption with respect to the mesh size $\Delta$. We observe again the expected first-order of convergence. The origin of this mass absorption has been previously analyzed and the results are in agreement with the "porosity" of the coupled VOF/embedded boundary method. Indeed, in a cell with three phases, the solid fraction is simply ignored (the solid is assumed to be ``porous'' at the $\Delta$ scale).

\subsection{2D sessile droplet on an embedded plane}\label{subsec:subsec2}
\begin{figure}[!h]
\centering
\includegraphics[trim={3cm 4cm 3cm 4cm},clip,width=0.8\linewidth]{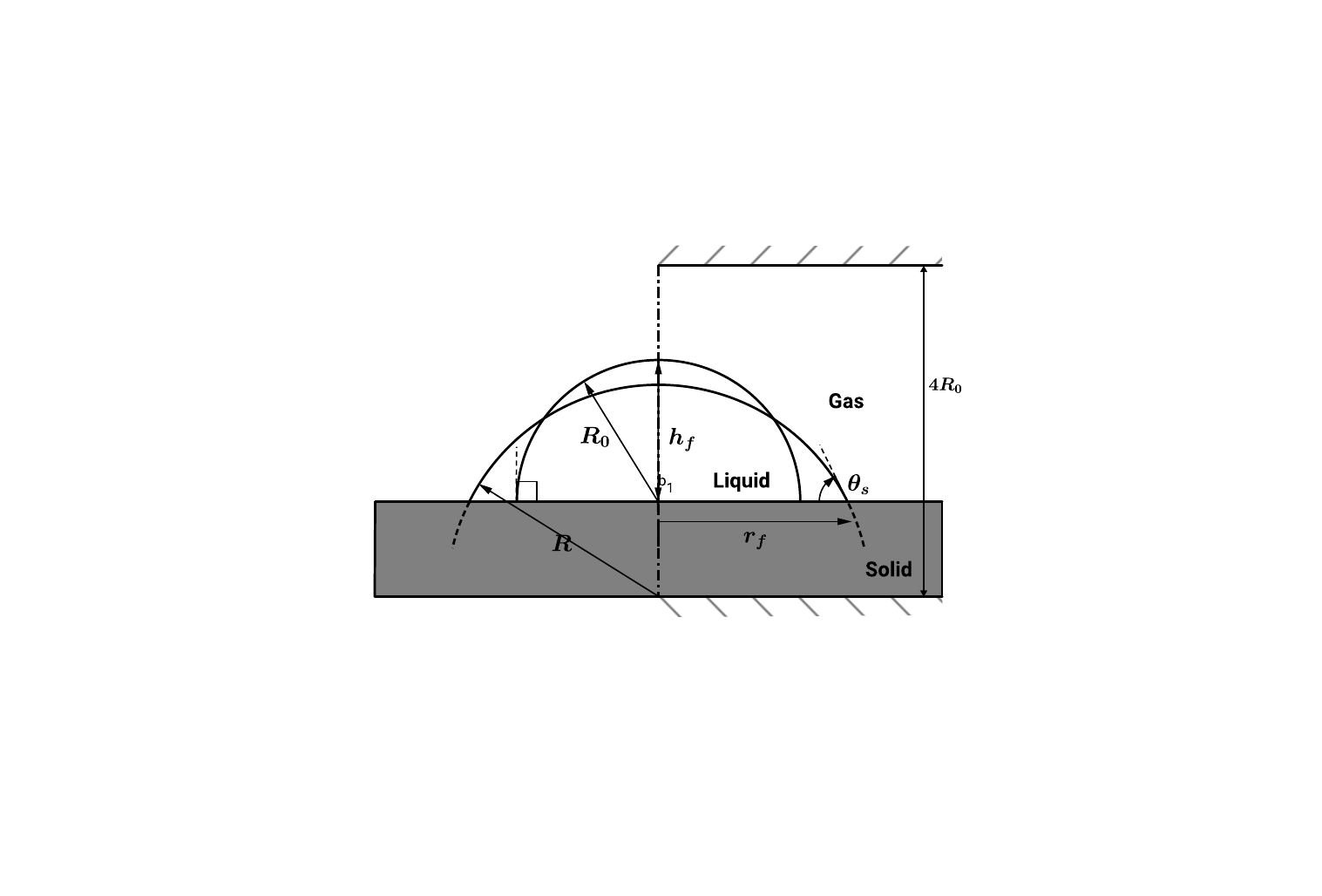}
\caption{Schematic representation of the initial and equilibrium shapes of a droplet on a flat surface with static contact angle $\theta_s$ and $Eo=0$.}
\label{fig:initial-config-sessile}
\end{figure}

In this test case, a two-dimensional drop of radius $R_0$ is released at rest with an initial contact angle $\theta_i=90^\circ$ on an embedded plane with a static contact angle $\theta_s$. Two possible orientations of the embedded plane are studied: either $0^\circ$ or $45^\circ$. In the first situation, we have an embedded geometry perfectly aligned with the grid, an ideal situation, whereas the latter configuration represents an embedded plane poorly aligned with the Cartesian grid.
Initially, the drop is placed in the domain as an half-disk (i.e. the initial contact angle is $90^\circ$, see Fig.\ref{fig:initial-config-sessile}). From its initial contact angle, the droplet will relax to reach its equilibrium position (the prescribed angle $\theta_s$).
If the effect of gravity is taken into account, the drop is also flattened on a horizontal plane or can move if the inclined plane is considered.
The Eötvös number $Eo$ is then used to quantify the relative effect of gravitational forces compared to surface tension forces: $Eo = \rho_LgR_0^2/\sigma$.\\

\subsubsection{Spreading droplet without gravity}
\begin{figure}[!h]
  \centering
  \begin{subfigure}[b]{0.4\textwidth}
  \includegraphics[width=1.3\textwidth]{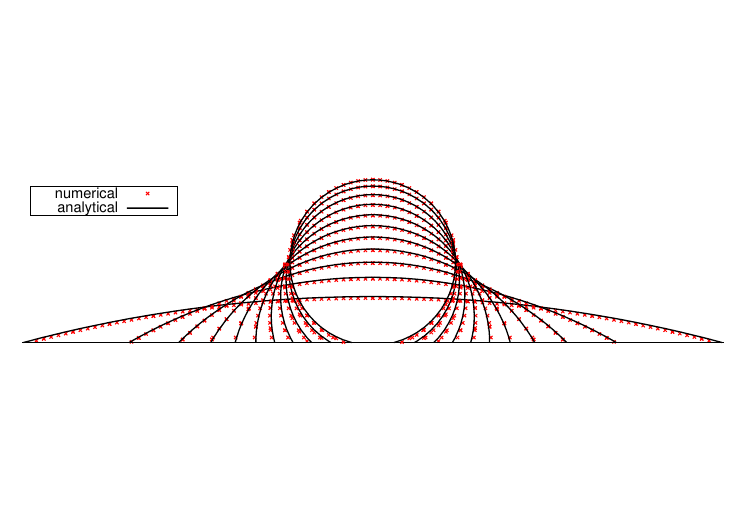}
  \caption{Horizontal plane}
    \label{fig:contact-angle-shape}
    \end{subfigure}
    \begin{subfigure}[b]{0.49\textwidth}
    \includegraphics[width=1.4\textwidth]{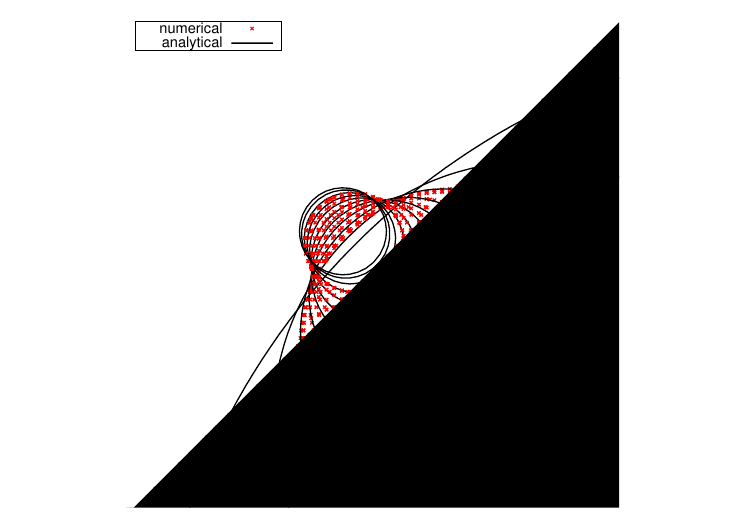}
    \caption{Inclined plane}
    \label{fig:contact-angle-shape-inclined}
    \end{subfigure}
    \caption{Comparison of the equilibrium droplet shapes against the analytical results (.) for $15^\circ<\theta_s<165^\circ$ without gravity $Eo=0$.}
\end{figure}

\begin{figure}[!h]
  \centering
  \begin{subfigure}[c]{0.49\textwidth}
  \includegraphics[width=1.2\textwidth]{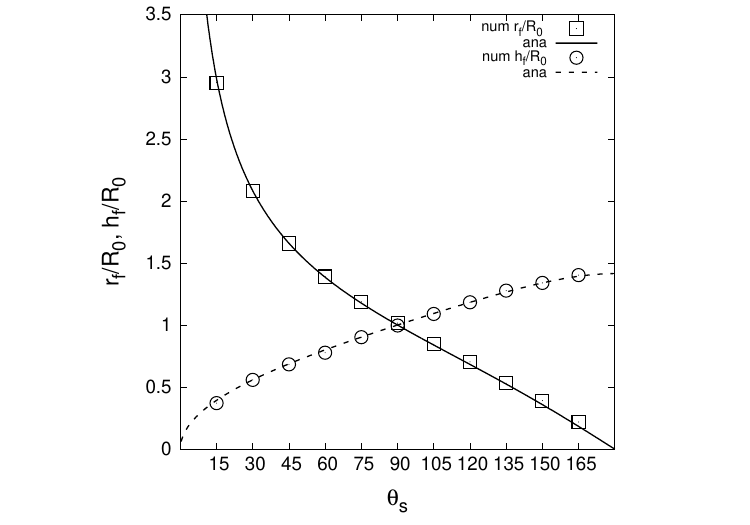}
  \caption{Horizontal plane}
    \label{fig:contact-radius}
    \end{subfigure}
    \begin{subfigure}[c]{0.49\textwidth}
    \includegraphics[width=1.2\textwidth]{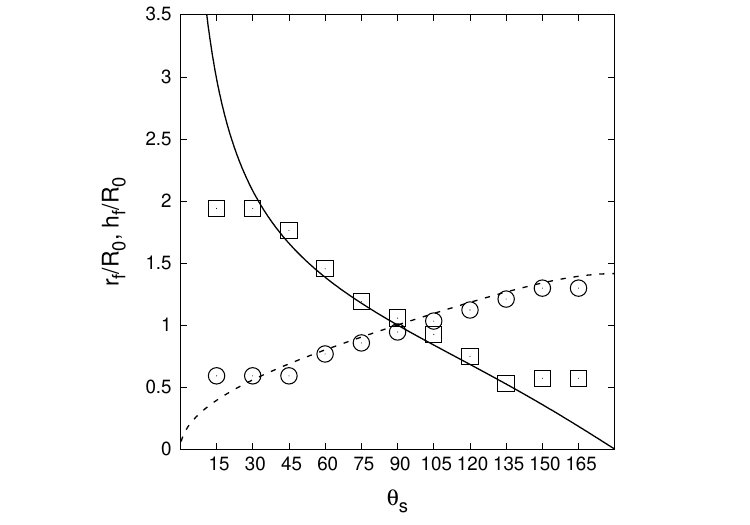}
    \caption{Inclined plane}
    \label{fig:contact-radius-inclined}
    \end{subfigure}
    \caption{Dimensionless droplet height $h_f/R_0$ and radius $r_f/R_0$ evolution for the equilibrium shape of a droplet on a flat surface without gravity $Eo=0$.}
\end{figure}

First, the simulation is performed for the case where gravity is zero ($Eo=0$). In the absence of gravity, the drop is hemispherical and its shape is controlled by surface tension only.
As the total drop volume $V$ is constant, it is possible to geometrically calculate at equilibrium the radius of the circle $R_f$, the radius of spreading $r_f$ and the height of the drop $h_f$.
\begin{subequations}
\begin{align}
R_f &=R_0\sqrt{\frac{\pi}{2(\theta_s-\sin\theta_s\cos\theta_s)}} \label{eq:Radius_contact}\\
h_f &=R_f(1-\cos\theta_s) \label{eq:height_contact}\\
r_f &=R_f\sin\theta_s \label{eq:radius_contact}
\end{align}
\end{subequations}
The simulation is still performed in a two dimensional box of size $2D\times2D$ with $D=2R_0$. All the boundaries are no slip walls and the contact angle boundary condition is applied as described above. The diameter $D$ is chosen as the characteristic length and set to $1$ in order to have $We\approx 1$ and $Re=1/\mu_L$. The adaptive mesh refinement is not activated here and the numerical domain is a Cartesian regular grid with the number of cells $N=64, 128, 256$ and $512$. Only $32$ cells describe the initial diameter $D$ of the drop for $N=64$ (and thus $R_0=N\Delta/4$ accordingly).
We study the case where the contact angle $\theta_s$ is varied between $15^\circ$ and $165^\circ$. The comparison between the numerical and the analytical equilibrium droplet shapes shows very satisfactory qualitative matching (Fig.~\ref{fig:contact-angle-shape}) for the horizontal embedded configuration. In the inclined plane situation, the numerical shape of the equilibrium droplet does not match perfectly the analytical one specially in the most hydrophobic or hydrophilic situation.
This issue is clearer when comparing the analytical droplet height $h_f$ and contact radius $r_f$ to the numerical results. Figure \ref{fig:contact-radius} presents the numerical droplet height $h_f$ and contact radius f$r_f$ for the equilibrium shape compared with the analytical results of Eqs.~\eqref{eq:height_contact}-\eqref{eq:radius_contact}, respectively as a function of $\theta_s$.
For the horizontal embedded configuration; all the considered contact angles show a good agreement with the analytical values. \\

It is less obvious for the inclined plane where both the numerical height $h_f/R_0$ and radius $r_f/R_0$ diverge from the analytical data for extreme angles ($15^\circ$, $30^\circ$  and $15^\circ$, $165^\circ$). This problem, already mentioned in the previous test case (see sec.~\ref{subsec:subsec1}), is linked to the poor grid alignment with the embedded boundary. With a $45^\circ$ inclined plane, the VOF interface might cross the embedded boundary and the Cartesian grid in such a way that the VOF flux from one cell to another is close to zero and thus the interface gets pinned. It is to note that the pinning of the contact line is perfectly symmetric between the hydrophobic and the hydrophilic case in that specific configuration ($45^\circ$).

\begin{figure}[!h]
  \centering
  \begin{subfigure}[c]{0.49\textwidth}
  \includegraphics[width=1.2\textwidth]{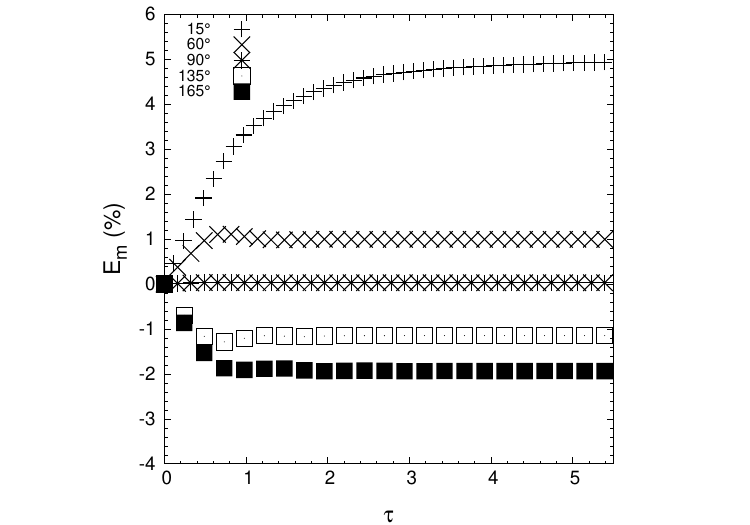}
  \caption{}
    \label{fig:sessile-error-mass-mesh128}
    \end{subfigure}
    \begin{subfigure}[c]{0.49\textwidth}
    \includegraphics[width=1.2\textwidth]{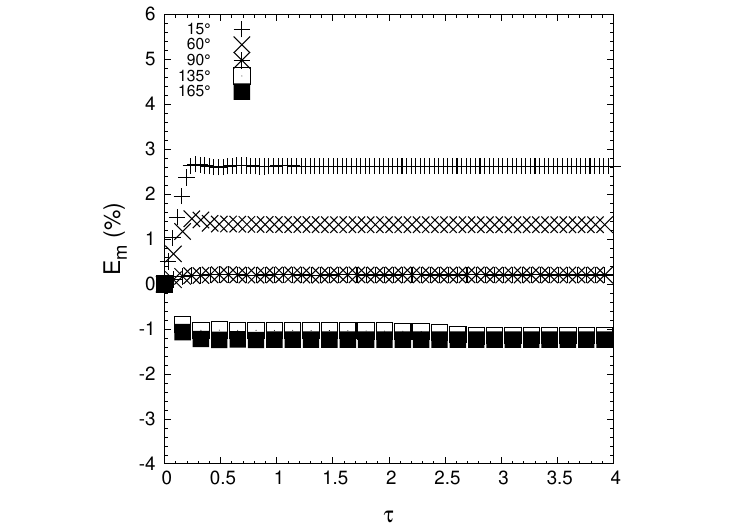}
    \caption{}
    \label{fig:sessile-inclined-error-mass-mesh64}
    \end{subfigure}
    \caption{Relative mass absorption for the sessile droplet, for different contact angles $\theta_s$ for $N=64$ for the two different orientations of the substrate: (a) Horizontal plane; (b) $45^{\circ}$ Inclined plane.}
\end{figure}

We also define a quantitative error by the mean difference of the volume fraction of the equilibrium drop shape and the volume fraction computed with the corresponding analytical shape:
\begin{equation}
E_F=  \frac{1}{N}\sum_{i=1}^{N}\left|({F_i^{ref}} - F_i)\right|
\label{eq:erreur-fraction}
\end{equation}
\begin{figure}[!h]
  \centering
  \begin{subfigure}[c]{0.49\textwidth}
  \includegraphics[width=1.2\textwidth]{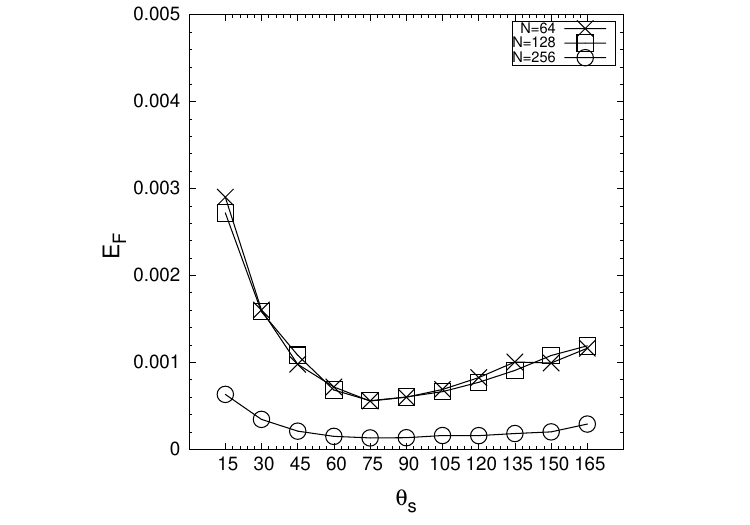}
  \caption{}
    \label{fig:errorshape}
    \end{subfigure}
    \begin{subfigure}[c]{0.49\textwidth}
    \includegraphics[width=1.2\textwidth]{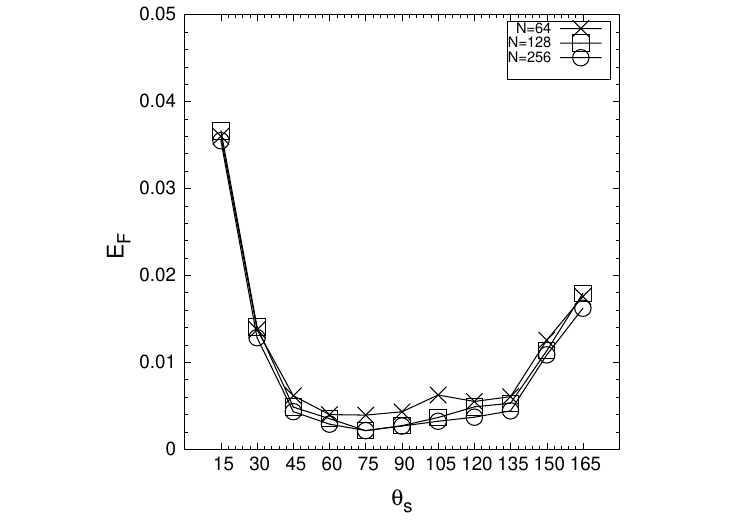}
    \caption{}
    \label{fig:errorshape-inclined}
    \end{subfigure}
    \caption{Average error on the volume fraction $E_F$ with different grids refinement, (a) Horizontal plane, (b) $45^{\circ}$ Inclined plane.}
\end{figure}
Figure \ref{fig:errorshape} and \ref{fig:errorshape-inclined} show the average error $E_F$ defined in equation \eqref{eq:erreur-fraction} for the horizontal and the inclined plane.
Surprisingly, a non-monotonic convergence of $E_F$ is observed with the grid refinement whatever the angle considered and the embedded configuration. In both situations, the maximum error is obtained for extreme angles ($15^\circ$ and $165^\circ$) with a higher level of error (by about a factor $10$) for the inclined plane though, mostly for the reasons set out above.\\
The mass absorption of the method is also examined and reported in Fig.\ref{fig:sessile-error-mass-mesh128} and Fig.\ref{fig:sessile-inclined-error-mass-mesh64} for the horizontal and $45^\circ$ inclined embedded plane and for different constant angles $\theta_s$.  \\
It appears that for the extreme angles namely $15^\circ$ and $165^\circ$ the maximum mass absorption is observed. As in the previous test case (see sec.~\ref{subsec:subsec1}), this is well quantified and linked to the ``porosity" of the coupled VOF/embedded boundary method.\\
The spurious currents evolution for the spreading droplet on the flat solid surface is also quantified through the capillary number defined as:
\begin{equation}
Ca_\text{max} = \frac{\mu_l\vv_{\infty}}{\sigma}
\label{eq:capillary_max}
\end{equation}
\begin{figure}[!h]
\centering
\includegraphics[width=0.6\textwidth]{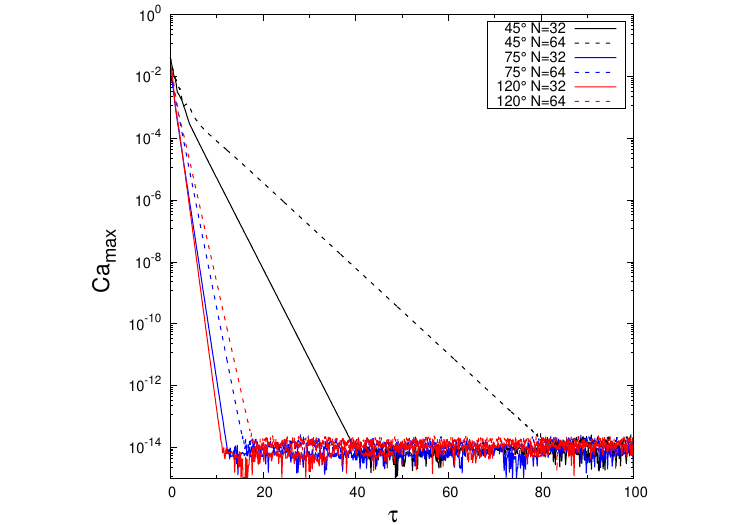}
\caption{Evolution of the maximum dimensionless velocity as the function of time for the spreading droplet on a horizontal surface with $\theta_s=45^\circ$, $\theta_s=75^\circ$  and $\theta_s=120^\circ$
and $N=32$ and $N=64$.
}
\label{fig:capillary_max}
\end{figure}
Figure \ref{fig:capillary_max} shows the evolution of the capillary number \eqref{eq:capillary_max} as a function of time for three different contact angle $\theta_s=45^\circ$,  $\theta_s=75^\circ$, $\theta_s=120^\circ$, two grid resolutions $N=32$ and $N=64$. For all combination of parameters, the maximum velocity eventually converges to zero within machine precision. 

\subsubsection{Influence of the gravity on a droplet resting on an embedded horizontal plane}
The influence of gravity ($Eo>0$) on the droplet shape is now presented.
While gravity tends to spread the drop, the surface tension keeps the drop as spherical as possible while the contact angle at the boundary must be satisfied.
For small Eötvös number ($Eo<<1$), the drop shape is controlled by surface tension and the maximum height of the drop $h_0$ is given by equation \eqref{eq:height_contact}.
At larger Eötvös number ($Eo>>1$), gravity influences the drop shape which starts to form a puddle. The maximum height $h_{\infty}$ of the drop is then proportional to the capillary length scale and given by the following expression \cite{DUPONT2010,asghar2023,PATEL2017}:
\begin{equation}
h_{\infty}=2\sqrt{\frac{\sigma}{\rho_Lg}}\sin\left(\frac{\theta_s}{2}\right)
\label{eq:height_contact_infty}
\end{equation}

\begin{figure}[!h]
\begin{subfigure}[t]{0.49\textwidth}
  \centering
  \includegraphics[width=\textwidth]{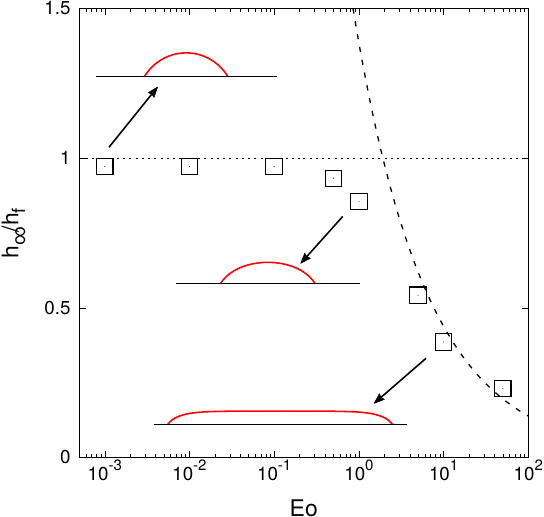}
  \caption{}
  \label{fig:contact-radius-Eotvos60}
  \end{subfigure}
  \begin{subfigure}[t]{0.49\textwidth}
  \centering
  \includegraphics[width=\textwidth]{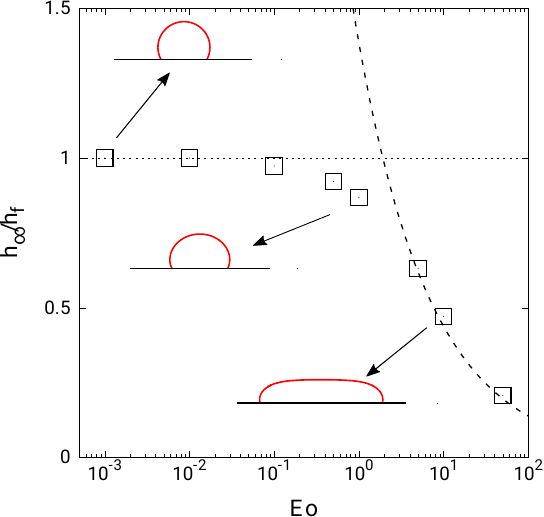}
  \caption{}
  \label{fig:contact-radius-Eotvos120}
  \end{subfigure}
  \caption{Dimensionless droplet height $h_{\infty}/h_f$ evolution for the equilibrium shape of a droplet on a flat surface for a varying Eötvös number: $10^{-3}<Eo<50$: (a) $\theta_s=60^\circ$, (b) $\theta_s=120^\circ$. {See also \url{http://basilisk.fr/sandbox/tavares/test_cases/sessile-embed-gravity.c}}}
\end{figure}
Figures \ref{fig:contact-radius-Eotvos60} and \ref{fig:contact-radius-Eotvos120} show the evolution of the dimensionless droplet height $h_{\infty}/h_f$ for different Eötvös numbers $10^{-3}<Eo<50$,  considering two contact angle values $\theta_s=60^\circ$ and $\theta_s=120^\circ$ and a variable Eötvös number $10^{-3}<Eo<50$. Around $Eo\approx 1$, we clearly notice the transition between the two regimes: the circular cap and the puddle. Overall, the numerical results shows satisfactory agreement with the two asymptotic solutions \eqref{eq:height_contact} and \eqref{eq:height_contact_infty}.

\subsection{3D sessile droplet on an embedded horizontal plane}\label{subsec:subsec3}
We investigate now the same spreading of a spherical droplet on an horizontal plane but in three dimensions.
A drop is initialized as a half-sphere of radius $R_0=0.25$ with  $\theta_i = 90^\circ$.
The analytical equilibrium radius of the droplet is given by:
\begin{equation}
R_0=\left(\frac{3V}{4\pi}\right)^{1/3}
\label{eq:radius3d}
\end{equation}
The contact angle is varied between $30^\circ$ and $150^\circ$. The drop oscillates and eventually relaxes to its equilibrium position $\theta_s$.
Here, the adaptive mesh refinement is enabled so that the maximum level of refinement corresponds to $R=16\Delta$.
\begin{figure}[!h]
  \centering
  \begin{subfigure}[c]{0.45\textwidth}
   \includegraphics[width=0.8\textwidth]{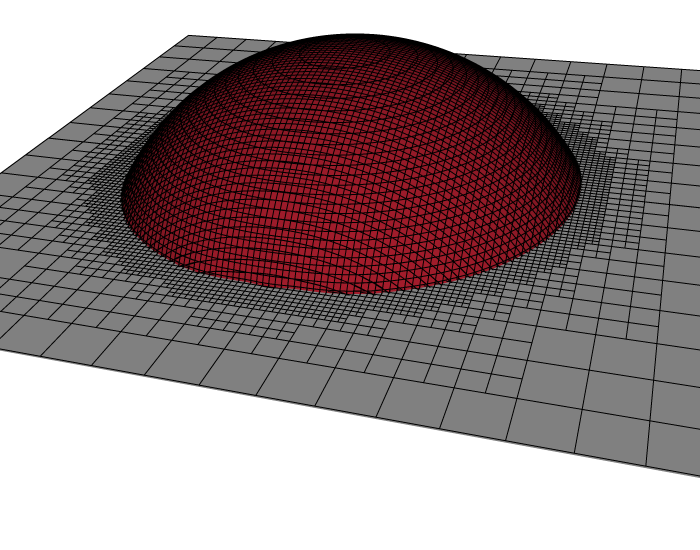}
     \caption{Equilibrium shape for $60^\circ$}
		\label{fig:contact-angle-shape-3D}
    \end{subfigure}
    \begin{subfigure}[c]{0.5\textwidth}
     \includegraphics[width=\textwidth]{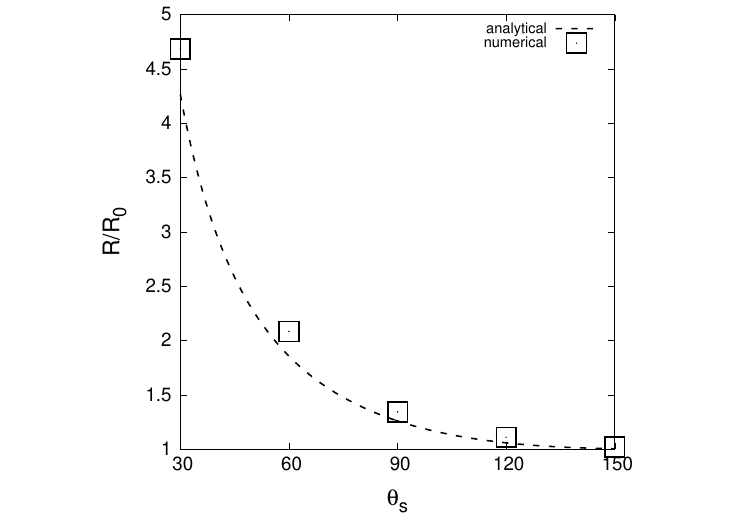}
     \caption{}
		 \label{fig:comp-ana-num-3D}
     \end{subfigure}
     \caption{(a) Equilibrium shape of 3D sessile droplet for $\theta_s = 30^\circ$, (b) Comparison of $ R/R_0$ to the analytical expression, with $R_0=(\frac{3V}{4\pi})^{1/3}$, See also \url{http://basilisk.fr/sandbox/tavares/test_cases/sessile-embed3D.c}}
\end{figure}
The comparison between the numerical and the analytical equilibrium droplet shapes shows satisfactory qualitative matching (Fig.~\ref{fig:comp-ana-num-3D}), even for such a coarse grid. This demonstrates that the algorithm developed for the contact angle boundary condition is also compatible with adaptive mesh refinement.

\subsection{Dynamics of the droplet on the embedded horizontal plane}
So far we have studied the asymptotic regime of a configuration to validate that our simulations converge numerically to the equilibrium state. No specific treatment was made for the dynamics and we simply relied on numerical relaxation.
A physically-correct description of the problem must involved two coupled numerical ingredients: the computation of the viscous stress in the mixed cell and the moving contact line.
In particular, we know that without any correction, the contact line dynamics should not converge since in the continuum limit, the no-slip boundary condition does not allow the contact line to move~\cite{Dussan1979}.

In the following test case, the dynamics of the spreading droplet is therefore studied.
The objective is to first emphasize the grid dependency when using a no-slip boundary condition on an embedded solid thus characterizing the effect of the ``numerical slip" and then introduce a Navier slip boundary condition Eq.~\eqref{eq:Navier} in order to (partially) regularize the singularity at the contact line.
The contact line behavior as a function of the boundary condition has been debated in many studies for the past decades but remains a numerical challenge especially in the context of hybrid method such as the VOF/embedded boundary in this work. 
Note that for the Navier boundary condition, grid convergence can be reached if the full hydrodynamic problem is solved in the slip length region \cite{Afkhami_2009,Legendre_2015,Fullana_2020}.
Here, we consider again the spreading of a semicircular droplet on an embedded solid boundary. We investigate the dynamics of the contact angle when the droplet relaxes to its equilibrium shape given by $\theta_s$.
The physical parameters are identical to those in section \ref{subsec:subsec2}.
The contact angle is fixed at $\theta_s=60^\circ$ and the mesh size is varied from $N=32$ to $N=256$. Gravity is ignored in this simulation.
Either a no-slip  ($\lambda=0$) or a Navier slip ($\lambda\ne 0$) condition is imposed along the embedded solid using the procedure described before Eqs.~\eqref{eq:dirichlet-gradient} and \eqref{eq:Navier-component}.
In this simulation, $\lambda$ is chosen as an adjustable parameter.
Figure \ref{subsec:subsec2} shows the evolution of the droplet radius $r$ with the dimensionless time for different values of $\lambda$ and a fixed grid $N=32$.
\begin{figure}[!h]
 \centering
  \includegraphics[width=0.6\linewidth]{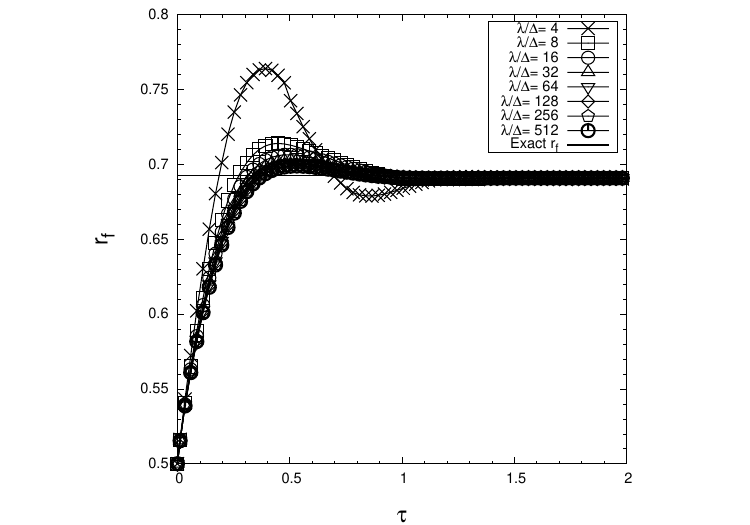}
  \caption{Evolution of the radius $r$ as a function of the dimensionless time $\tau$ for different values of $\lambda$, $N=32$.}
  \label{fig:Contact_ray_lambda}
\end{figure}
For all values of $\lambda$, a convergence of the droplet radius $r$ is finally reached. However, as the resolution of $\lambda$ is increased, we observe a quick convergence of the radius. At least $4$ cells are required to describe the slip length $\lambda$ and a convergence is reached when $\lambda=16\Delta$ for $N=32$. This value of $\lambda=16\Delta_{32}$ is therefore adopted for the rest of the simulations and for all the grids.

\begin{figure}[!h]
\begin{subfigure}[t]{0.49\textwidth}
 \centering
  \includegraphics[trim={0cm 0cm 0cm 0cm},clip,width=\linewidth]{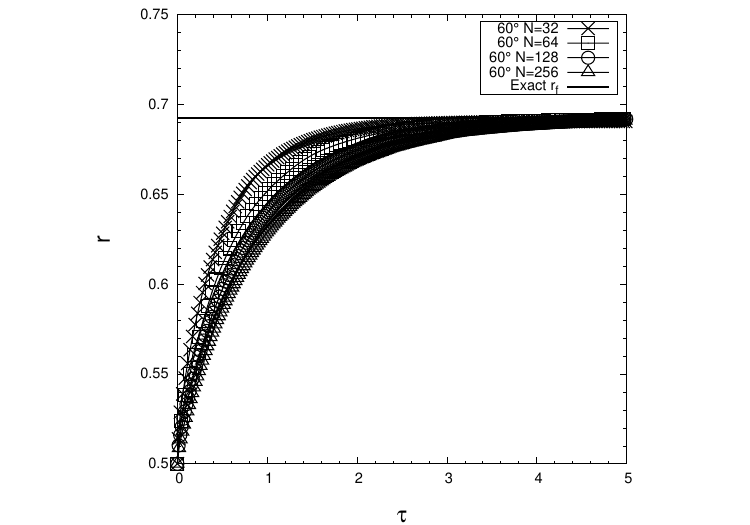}
  \caption{}
  \label{fig:comp-navier0}
\end{subfigure}
\begin{subfigure}[t]{0.49\textwidth}
 \centering
  \includegraphics[trim={0cm 0cm 0cm 0cm},clip,width=\linewidth]{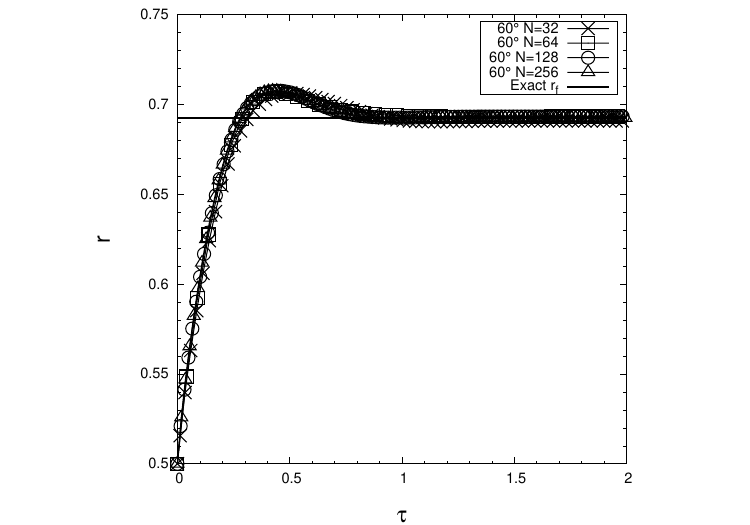}
  \caption{}
  \label{fig:comp-navier0.75}
 \end{subfigure}
\begin{subfigure}[t]{0.49\textwidth}
 \centering
  \includegraphics[trim={0cm 0cm 0cm 0cm},clip,width=\linewidth]{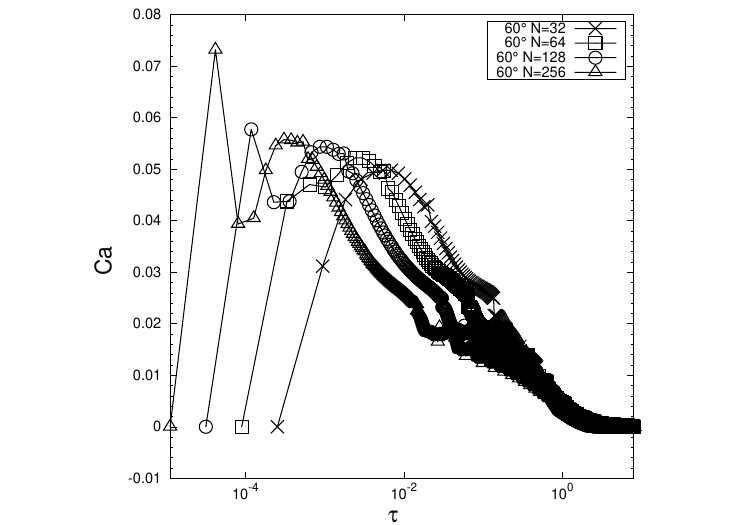}
  \caption{}
  \label{fig:capillary-navier0}
\end{subfigure}
\begin{subfigure}[t]{0.49\textwidth}
 \centering
  \includegraphics[trim={0cm 0cm 0cm 0cm},clip,width=\linewidth]{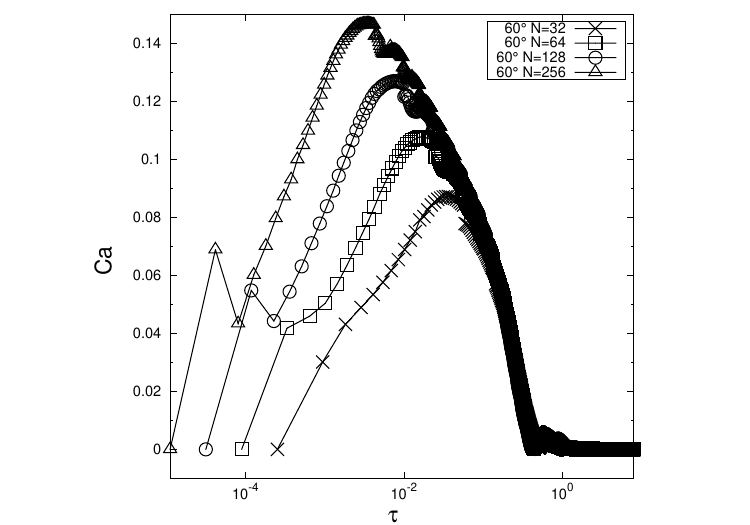}
  \caption{}
  \label{fig:capillary-navier0.75}
 \end{subfigure}
\begin{subfigure}[t]{0.49\textwidth}
 \centering
  \includegraphics[trim={0cm 0cm 0cm 0cm},clip,width=\linewidth]{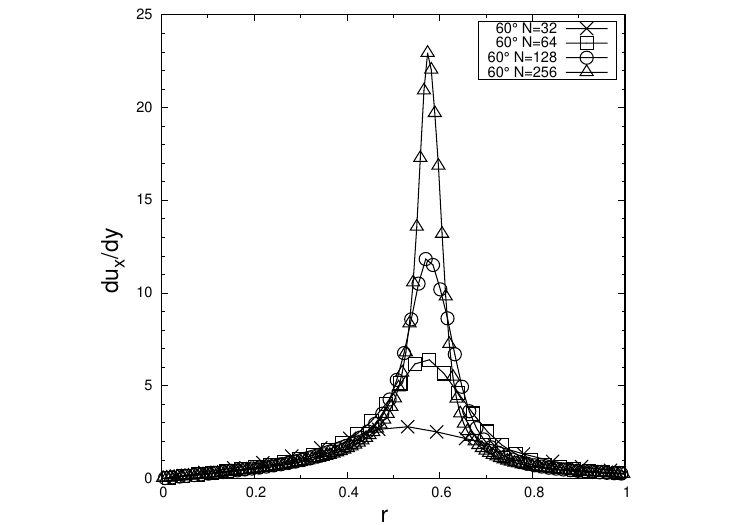}
  \caption{}
  \label{fig:gradient-navier0}
\end{subfigure}
\begin{subfigure}[t]{0.49\textwidth}
 \centering
  \includegraphics[trim={0cm 0cm 0cm 0cm},clip,width=\linewidth]{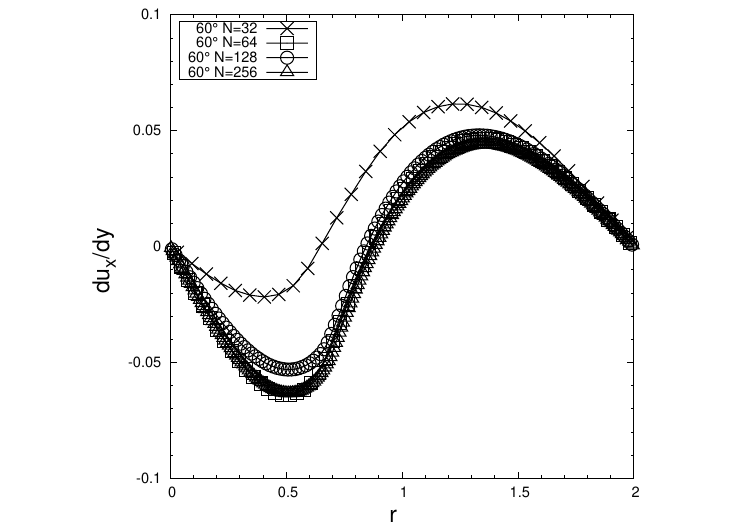}
  \caption{}
  \label{fig:gradient-navier0.75}
 \end{subfigure}
 \caption{Grid convergence using the no slip boundary condition $\lambda=0$ (left) and the Navier slip boundary condition $\lambda=16\Delta$ (right): Evolution of the droplet radius $r$ as a function of the dimensionless time $\tau$ (a) no slip, (b) Navier slip, Capillary number evolution (c) no slip, (d) Navier slip, Velocity gradient $d\vv/dn$ at $\tau=0.5 $ (e) no slip, (f) Navier slip.}
 \end{figure}

Considering now the grid convergence, we compare both the no-slip case and the Navier slip case.
In both cases, the radius converges to the terminal radius $r_f$.
As expected, the radius $r$ does not converge with mesh refinement fig.~\ref{fig:comp-navier0} in the no-slip case.
The contact line velocity decreases as the mesh resolution is increased.
It is a consequence of the numerical slip length which is larger for coarser meshes: as the slip decreases with finer meshes, so does the contact line velocity.
On the other hand, the use of the Navier slip boundary condition changes the spreading dynamics of the droplet.
By introducing a Navier slip length larger than the grid cell, the results tends to be independent from mesh refinement. Moreover a faster drop spreading is observed in that situation.
The Navier boundary condition coupled with a constant contact angle is then sufficient to achieve grid convergence. 

\subsection{Drop impact on a fiber}
An additional validation study for a 3D drop impact on a circular fiber is presented here. The parameters used in the simulations are inspired by the experiment of \cite{Lorenceau_2004} and the related numerical simulation of \cite{Wang_2018} where they studied a liquid droplet impacting a thin solid fiber in a gaseous environment. In both works, the static contact angle was chosen to be $10^\circ$ and two different impact velocities were studied. 
\begin{table}[!h]
\centering
\begin{adjustbox}{max width=\textwidth}
\begin{tabular}{cccccc}
\hline
 & $\rho_l/\rho_g$   & $\mu_l/\mu_g$ & $Re$ &  $We$ & $Fr$ \\
\hline
\hline
Case 1 & 1000  & 50 & 175 & 1.6 & 2.13 \\
Case 2 & 1000  & 50 & 830 & 36.85 & 10 \\
\hline
\end{tabular}
\end{adjustbox}
\caption{Non dimensional parameters for the 3D drop impacting a circular fiber.}
\label{tab:adimensional-parameters}
\end{table}
Here, the drop has a diameter of $D$ and an impact velocity of $U$, chosen as the characteristic length and velocity of the problem. The circular fiber has a diameter $Df$ so that the ratio $Df/D\approx0.485$, which is slightly larger than the reference fiber diameter ($Df/D\approx0.416$ see \cite{Wang_2018}). 
We consider two cases where the impact velocity is varied so that we get the non-dimensional parameters, $Re=\frac{\rho_l U D}{\mu_l}$ , $We=\frac{\rho_l U^2 D}{\sigma}$, $Fr=\frac{U^2}{g D}$, given in table.~\ref{tab:adimensional-parameters}. We chose here to work with a static contact angle $\theta_s=15^\circ$.
The domain size is ${4D}^3$ discretized with an uniform mesh of $128^3$, resulting in $Df/\Delta \approx 14.7$. 
\begin{figure}[!h]
\begin{subfigure}[t]{0.19\textwidth}
\includegraphics[trim={15cm 10cm 2cm 2cm},clip,width=2.7\textwidth]{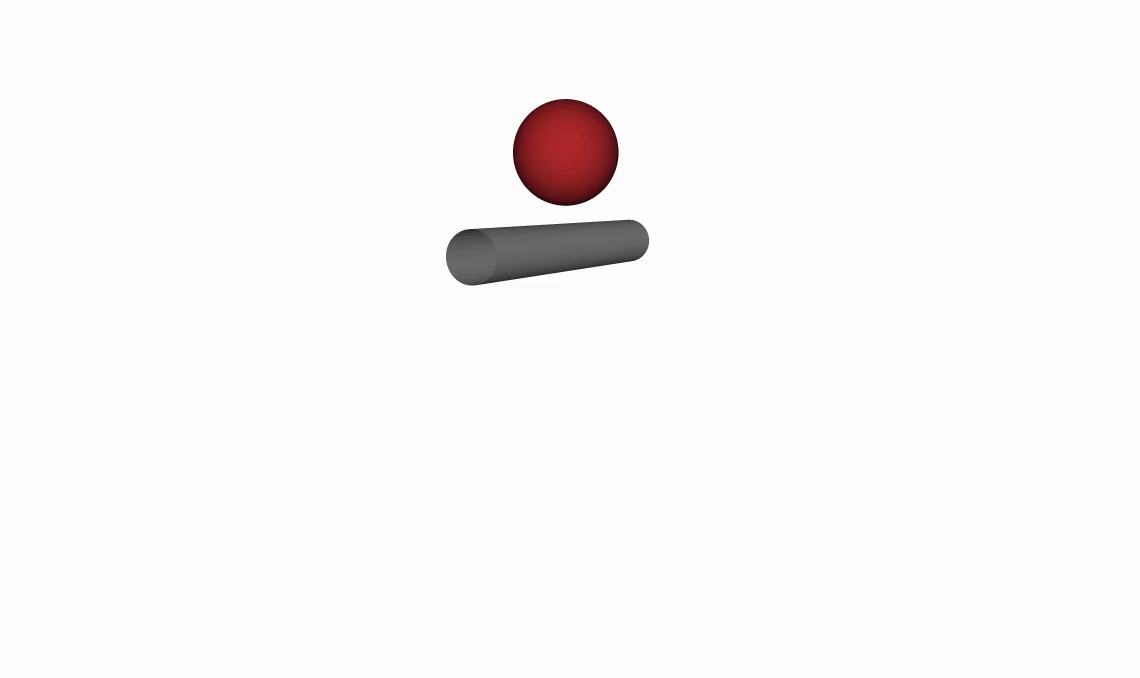}
\end{subfigure}
\begin{subfigure}[t]{0.19\textwidth}
\includegraphics[trim={15cm 10cm 2cm 2cm},clip,width=2.7\textwidth]{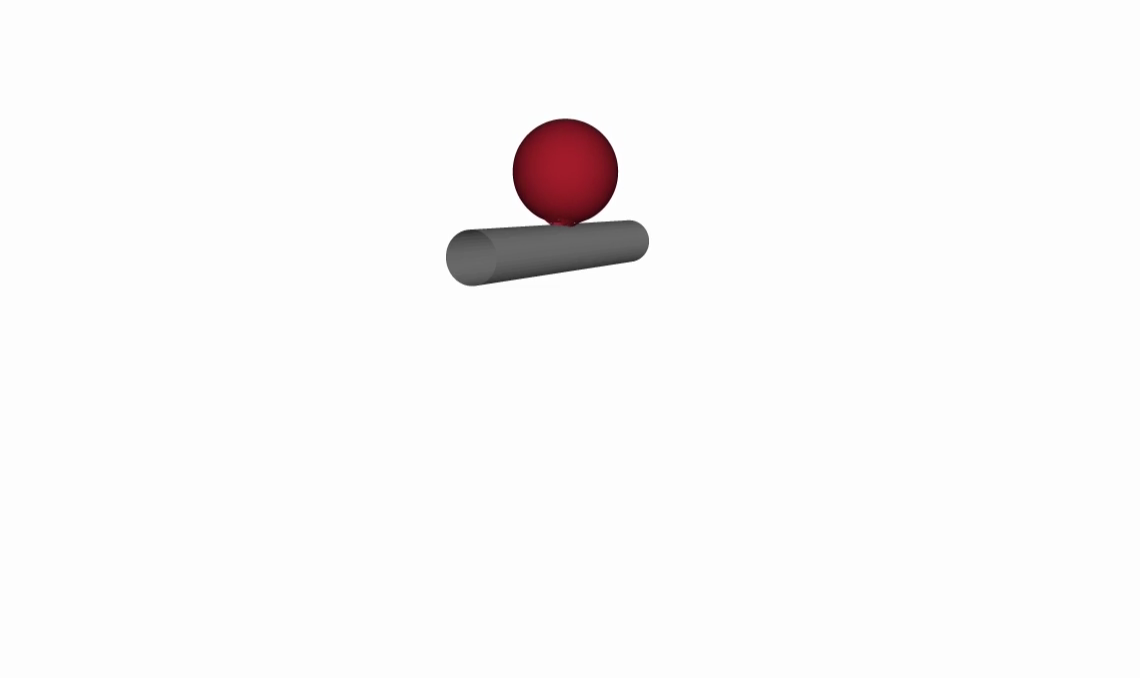}
\end{subfigure}
\begin{subfigure}[t]{0.19\textwidth}
\includegraphics[trim={15cm 10cm 2cm 2cm},clip,width=2.7\textwidth]{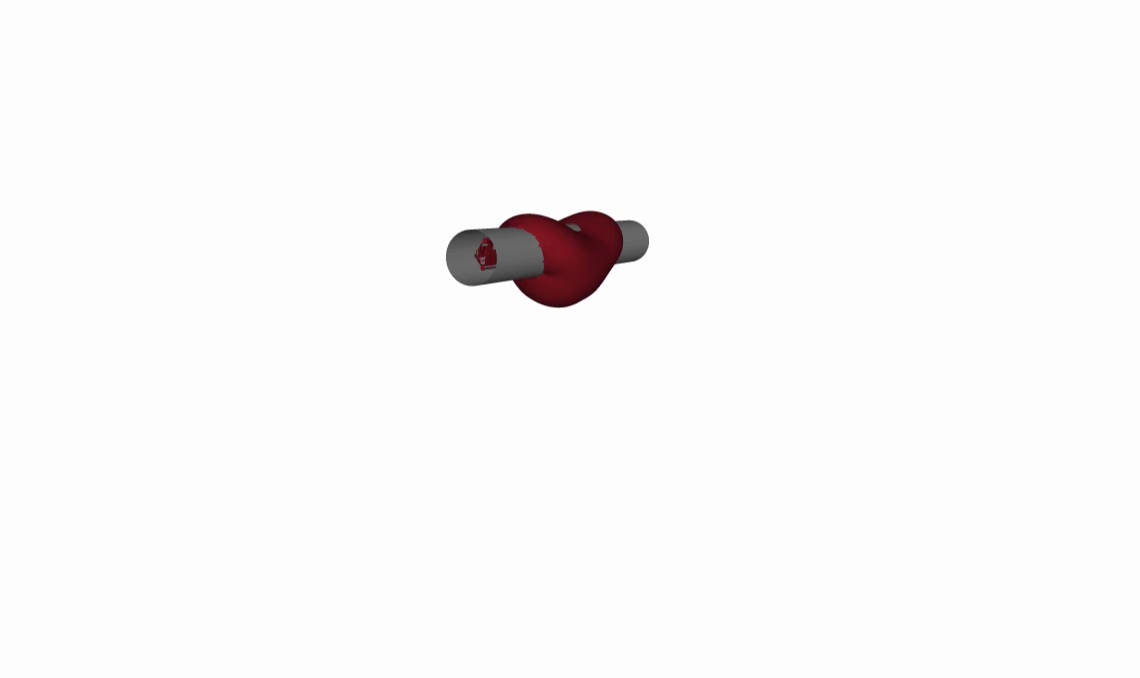}
\end{subfigure}
\begin{subfigure}[t]{0.19\textwidth}
\includegraphics[trim={15cm 10cm 2cm 2cm},clip,width=2.7\textwidth]{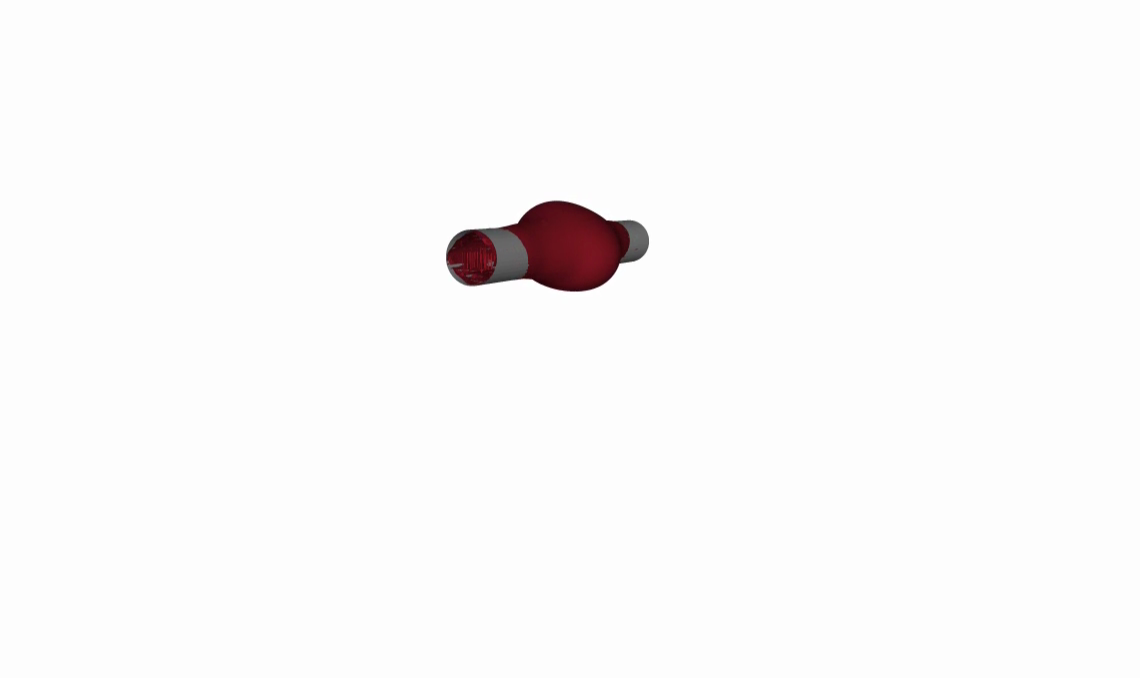}
\end{subfigure}
\begin{subfigure}[t]{0.19\textwidth}
\includegraphics[trim={15cm 10cm 2cm 2cm},clip,width=2.7\textwidth]{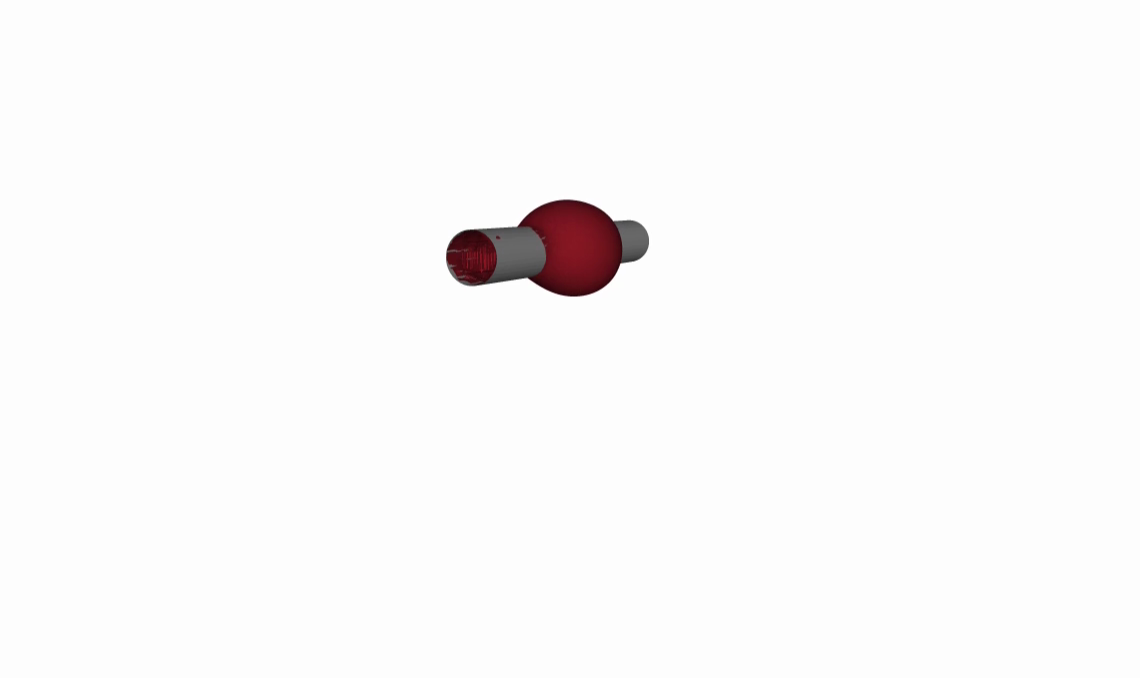}
\end{subfigure}
\caption{A drop falling at $Re=175$ and $\theta_s=15^\circ$ is captured by the fiber.}
\label{fig:drop-captured}
\end{figure}

\begin{figure}[!h]
\begin{subfigure}[t]{0.19\textwidth}
\includegraphics[trim={15cm 6cm 2cm 3cm},clip,width=2.7\textwidth]{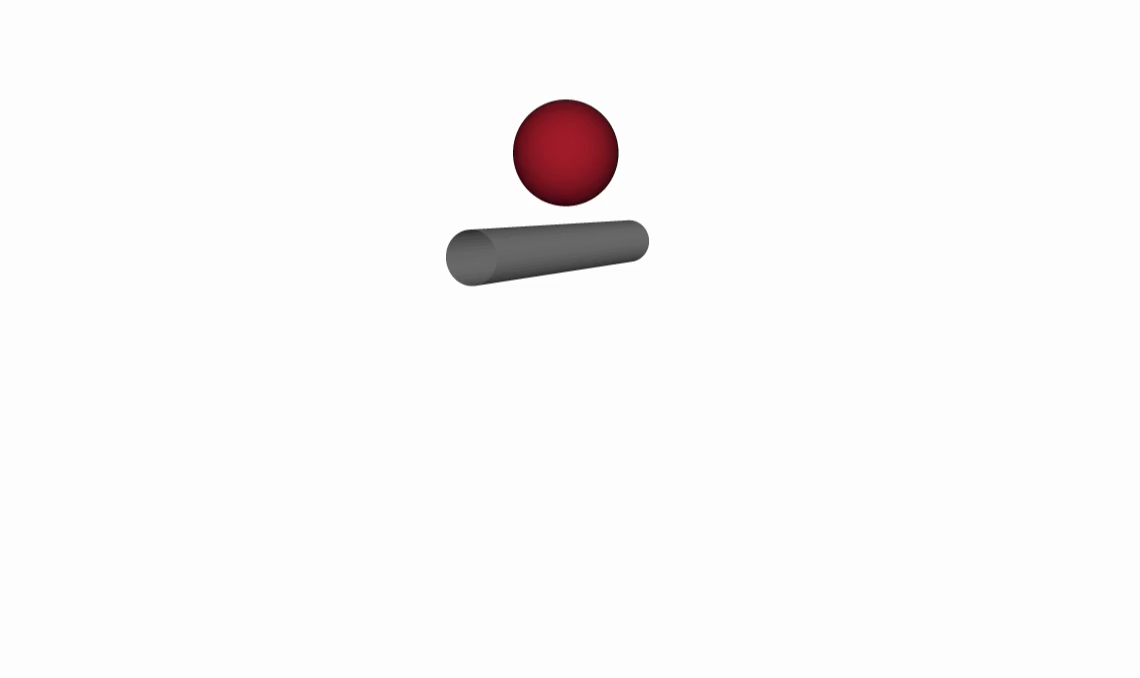}
\end{subfigure}
\begin{subfigure}[t]{0.19\textwidth}
\includegraphics[trim={15cm 6cm 2cm 3cm},clip,width=2.7\textwidth]{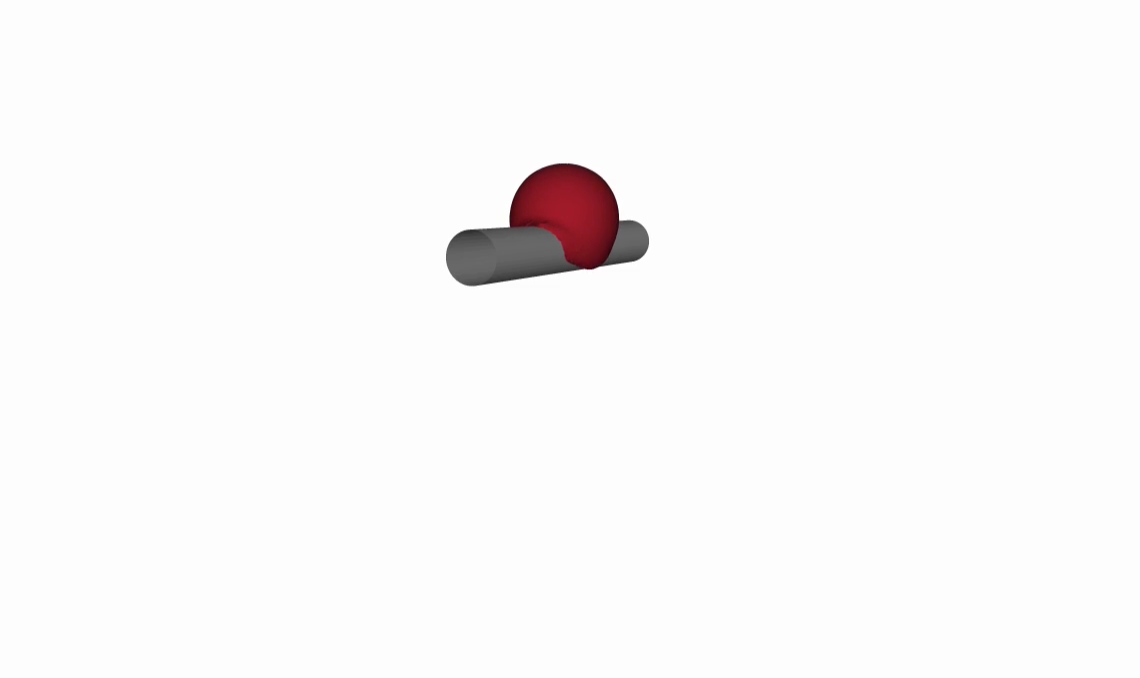}
\end{subfigure}
\begin{subfigure}[t]{0.19\textwidth}
\includegraphics[trim={15cm 6cm 2cm 3cm},clip,width=2.7\textwidth]{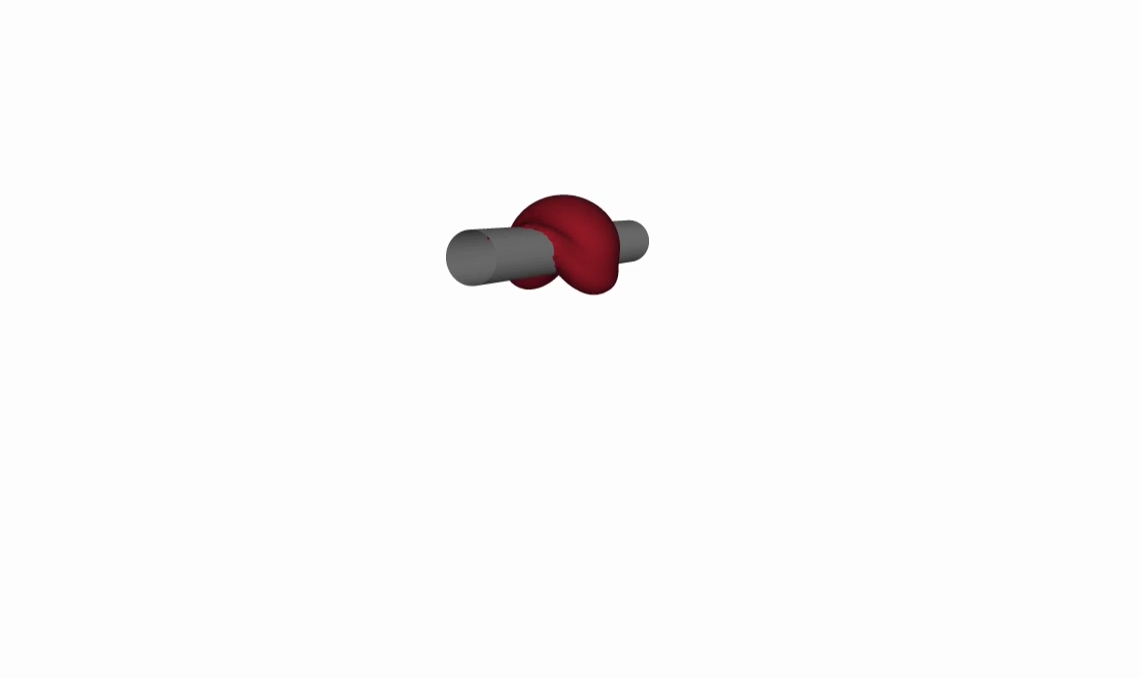}
\end{subfigure}
\begin{subfigure}[t]{0.19\textwidth}
\includegraphics[trim={15cm 6cm 2cm 3cm},clip,width=2.7\textwidth]{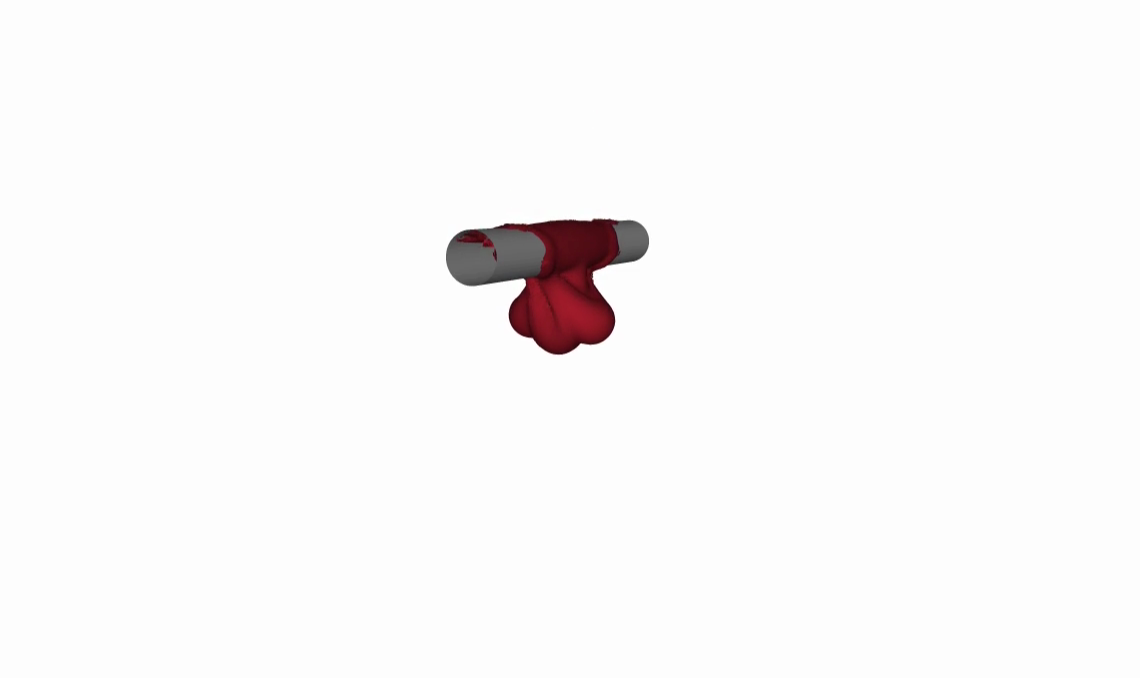}
\end{subfigure}
\begin{subfigure}[t]{0.19\textwidth}
\includegraphics[trim={15cm 6cm 2cm 3cm},clip,width=2.7\textwidth]{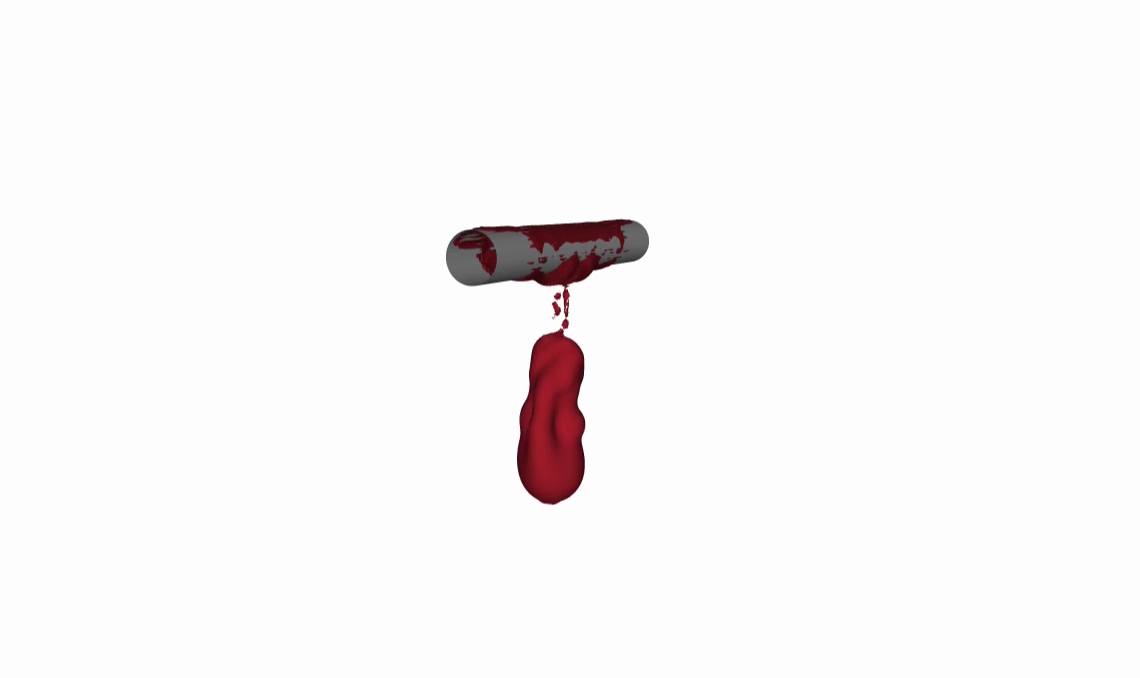}
\end{subfigure}
\caption{A drop falling at $Re=830$ and $\theta_s=15^\circ$ detaches from the fiber.}
\label{fig:drop-detached}
\end{figure}
Figures \ref{fig:drop-captured} and \ref{fig:drop-detached} show snapshots of impact dynamics on a fiber. 
For the first case, the drop is captured by the fiber whereas the drop detaches from the fiber in the second case as in the experimental and numerical references.
Our 3D simulations show good agreement with both the experimental and numerical observations of the aforementioned references. We do not perform here a more quantitative validation since it would require a better characterization of the experimental conditions, which is beyond the scope of this work.

\section{Conclusion and discussion}
In this work, a hybrid VOF/embedded boundary for the modelling of two-phase flows interacting with arbitrary solid geometries is presented.
A conservative second-order embedded boundary method is used here to simulate complex solid shapes while preserving the accuracy at the fluid-solid interface for Dirichlet/Neumann like boundary conditions. A specific effort has been devoted to propose a second-order accurate Navier/free slip boundary condition compatible with the embedded boundary method. Thanks to the geometric Volume Of Fluid method (VOF), the fluid-fluid interface is tracked sharply in a conservative way. 

To account for the contact angle boundary condition at the intersection between the solid and the fluid-fluid interface (triple point/line), an algorithm using fluid ``ghost cells'' within the solid region is proposed, allowing to reproduce contact line dynamics in the vicinity of the triple point/line.

Several test cases have been investigated to validate our method, namely the spreading of a droplet in various configurations (cylinder, horizontal or tilted plane, with or without gravity). In the majority of cases, a very satisfactory agreement with reference solutions is observed. The 3D spreading droplet case using Adaptive Mesh Refinement (AMR) also shows satisfactory results overlapping with the analytical solution.

For some configurations (tilted plane, cylinder) and for some values of the contact angle, a pinning of the contact line has been identified.
The alignment of the Cartesian grid with the solid leads in certain situations to a zero VOF flux advection resulting in a pinning of the contact line. This is an inherent problem of the method but one could imagine different solutions to overcome pinning situations, for instance imposing a slip length at the solid to enforce the motion of the contact line.

The dynamics of the spreading droplet have also been studied using either Dirichlet or Navier slip boundary conditions on the embedded boundary. Even if a time convergence has been observed in both configurations, a significant difference on the grid convergence is emphasized, as shown in previous studies in the literature. We showed that grid convergence can be obtained only using a Navier slip condition as long as the numerical slip length is well resolved by the grid.

A perspective of this study would be to test different contact angle models in the literature since the proposed algorithm is also compatible with a contact angle varying in space and time. A limitation, common also in other methods, is the case of ``shallow'' angles ($\theta_s<10^\circ$ or $\theta_s<170^\circ$) which would require special treatment.

The implementation of the numerical methods and test cases are freely accessible as part of the Basilisk open-source library (see \url{http://basilisk.fr/sandbox/tavares/test_cases/} ).\\

{\it Acknowledgement:} this work was partially supported by Agence de l'Innovation de D\'efense (AID) - via Centre Interdisciplinaire d'Etudes pour la D\'efense et la S\'ecurit\'e (CIEDS) - (project 2021 - ICING)
\newpage
\bibliographystyle{plain}
\bibliography{biblio}
\end{document}